\documentclass[twocolumn, prc, amssymb, superscriptaddress,aps,preprintnumbers,amsmath,floatfix]{revtex4}
\usepackage{multirow}
\usepackage{graphicx}% Include figure files
\usepackage{dcolumn}% Align table columns on decimal point
\usepackage{bm}% bold math
\usepackage{color}
\usepackage{amsmath}%[fleqn]
\usepackage{wasysym}
\usepackage{float}
\usepackage{array,url,kantlipsum}
\usepackage{verbatim}
\usepackage{xcolor}
\usepackage{siunitx}
\usepackage{CJK}
\usepackage{dsfont}
\usepackage{textcomp}
\newcolumntype{P}[1]{>{\centering\arraybackslash}p{#1}}
\usepackage{array}
\usepackage{physics}
\usepackage{braket}
\usepackage{lipsum}
\usepackage{booktabs}
\usepackage{upgreek}
\usepackage{chemformula}
\setchemformula{kroeger-vink}

\newcommand{\lrb}[1]{\left(#1 \right)} % saves time, lrb stands for left right bracket () 
\newcommand{\lrsb}[1]{\left [#1 \right]} %lrsb stands for left right squared bracket [] 
\newcommand{\lrcb}[1]{\left \{#1 \right \}}
\newcommand{\ab}[1]{\left |#1 \right |}
\begin{document}

%\title{Nuclear photoabsorption in $^{229}$Th using twisted light}

\title{Electronic Bridge processes in $^{229}$Th-doped LiCAF and LiSAF}

%%%%%%%%%%%%%%%%%%%%%%%%%%%%%%%%%%%%%%%%%%%%%%%%%%%%%%%%%%%%%%%%%%%%%%%%%%%%%%%%%

\author{Tobias \surname{Kirschbaum}}
\email{tobias.kirschbaum@uni-wuerzburg.de}
\affiliation{University of W\"urzburg, Institute of Theoretical Physics and Astrophysics,  Am Hubland, 97074 W\"urzburg, Germany}

\author{Martin \surname{Pimon}}
\affiliation{Institute for Theoretical Physics, TU Wien, Stadionallee 2, 1020 Vienna, Austria}

\author{Andreas \surname{Grüneis}}
\affiliation{Institute for Theoretical Physics, TU Wien, Stadionallee 2, 1020 Vienna, Austria}

\author{Thorsten \surname{Schumm}}
\affiliation{Institute for Atomic and Subatomic Physics, TU Wien, Stadionallee 2, 1020 Vienna, Austria}

%\email{tobias.kirschbaum@uni-wuerzburg.de}

\author{Adriana P\'alffy}
\email{adriana.palffy-buss@uni-wuerzburg.de}
\affiliation{University of W\"urzburg, Institute of Theoretical Physics and Astrophysics,  Am Hubland, 97074 W\"urzburg, Germany}

%%%%%%%%%%%%%%%%%%%%%%%%%%%%%%%%%%%%%%%%%%%%%%%%%%%%%%%%%%%%%%%%%%%%%%%%%%%%%%%%%

\date{\today}

\begin{abstract}

Electronic bridge mechanisms driving the \textsuperscript{229}Th  nuclear clock transition  in the vacuum-ultraviolet-transparent crystals \textsuperscript{229}Th:LiCAF (LiCaAlF$_6$) and \textsuperscript{229}Th:LiSAF (LiSrAlF$_6$) are investigated theoretically. Due to doping-induced symmetry breaking within the host crystal, electronic defect states emerge around the thorium nucleus and can facilitate nuclear (de)excitation via laser-assisted electronic bridge mechanisms. We investigate spontaneous and laser-assisted electronic bridge schemes  for different charge compensation mechanisms. 
While the calculated spontaneous electronic bridge rates are very small, laser-assisted electronic bridge schemes for nuclear (de)excitation turn out to be significantly more efficient than both spontaneous nuclear decay and direct laser excitation, offering  promising prospects for the future clock operation.

\end{abstract}
%%%%%%%%%%%%%%%%%%%%%%%%%%%%%%%%%%%%%%%%%%%%%%%%%%%%%%%%%%%%%%%%%%%%%%%%%%%%%%%%%

\maketitle

\section{Introduction}
\label{intro}
The $^{229}$Th nucleus is a unique candidate for the design and construction of the first nuclear clock \cite{peik2003nuclear, peik2015nuclear, peik2021nuclear,thirolf2024thorium}. From a nuclear physics point of view,  its first excited state at 8.36 eV is an isomer, i.e., a long-lived state, with a very narrow width [radiative lifetime of $\mathcal{O}(\SI{e3}{\per \second})$]. %This unique very low
This exceptionally low transition energy  renders the isomer  accessible to
 narrowband vacuum ultraviolet (VUV) light \cite{kraemer2023observation,zhang2024dawn}. A frequency standard operating on this nuclear transition would be insensitive to many shifts plaguing atomic transitions \cite{peik2021nuclear,dzuba2025nuclear}. Apart from several compelling technological applications,  an improved clock precision and in particular a nuclear clock would also allow the investigation of the temporal variation of fundamental constants setting the scales of the electromagnetic force and, as a new aspect, also of the strong force  \cite{flambaum_2006,Fadeev_2020,beeks2024fine}. 
In addition, it has been shown that the $^{229}$Th nuclear clock could potentially enhance the sensitivity for the detection of light dark matter \cite{peik2021nuclear,tsai2023direct,darkmatter25}.

At present, two practical approaches for the thorium nuclear clock are pursued. The first approach involves a single thorium ion stored in a trap \cite{peik2003nuclear,campbell2012single}. This setup promises outstanding accuracy \cite{campbell2012single}, though its realization is technically demanding due to the very low excitation rate per nucleus, and the required ultra-high vacuum and laser cooling. The second approach envisages a solid-state nuclear clock operating in thorium-doped VUV transparent crystals \cite{peik2003nuclear,rellergert2010,kazakov2012performance}. Here, a large number of dopants $N\sim 10^{14}$ up to $10^{16}$ can be interrogated simultaneously. This would be advantageous for the clock stability, which is proportional to $\sqrt{N}$ \cite{itano93, ludlow2015optical}. However, line broadening and nuclear level shifts and splittings induced by the intrinsic electromagnetic fields of the host crystal reduce the achievable accuracy \cite{kazakov2012performance,dessovic2014229thorium,pimon2022ab}. Also, due to the difference in valence states between thorium and the host crystal constituents, the doping process produces defects in the crystal and leads to the emergence of ``defect'' electronic states in the band gap. Various defect arrangements can occur, involving interstitial atoms, vacancies, or substitutional atoms situated around thorium, leading to distinct stoichiometric configurations known as charge compensation schemes, which may directly affect the clock operation and transition frequencies.

The solid-state approach has facilitated the recent experimental progress in driving and detecting the nuclear clock transition, despite surmounting the previously insufficient precision in the knowledge of the transition frequency, and VUV laser technology still being in development. The radiative decay of the nuclear excited state in VUV-transparent \ch{CaF_2}  crystals in which Th ions were implanted was reported in 2023~\cite{kraemer2023observation}, followed by successful laser excitation in \ch{Th:CaF_2} \cite{tiedau2024laser, zhang2024dawn, schaden2024laser}, \ch{Th:LiSAF} \cite{elwell2024laser,terhune2024photo}, and \ch{ThF_4} \cite{229thf4ThinFiZhang2024} in 2024. As of today  (2025), the nuclear isomer's radiative decay has been probed at various temperatures~\cite{higgins2025} and across multiple host materials, including bulk \ch{Th:CaF_2}, \ch{Th:MgF_2}, and \ch{Th:LiSAF}~\cite{pineda2025radiative}. Not all the reported observations have, at present, a clear explanation. As notable examples, there appears to be missing VUV fluorescence in several experiments \cite{kraemer2023observation, pineda2025radiative}, and laser spectra have shown additional narrow lines 
\cite{zhang2024dawn}. Also,  laser-induced quenching of the nuclear transition of unknown origin has been reported \cite{schaden2024laser, hiraki2024controlling, terhune2024photo}.

While it is assumed in Ref.~\cite{zhang2024dawn} that different charge compensation mechanisms in \ch{Th:CaF_2}  are responsible for the additional lines in the excitation spectrum, electronic processes in the crystal seem to be responsible for the other observed phenomena. One of the electronic process candidates potentially capable of modifying the nuclear decay rates is the Electronic Bridge (EB). EB is a third-order perturbative process in which the nucleus is (de)excited via electromagnetic coupling to the atomic shell, without requiring exact energy matching between electronic and nuclear transition energies. The energy mismatch is compensated by the emission or absorption of a photon. In the context of $^{229}$Th, a variety of EB scenarios involving thorium ions have been investigated theoretically \cite{strizhov1991decay,tkalya1992probability,porsev2010effect,bilous2018laser,bilous35p,porsev2021low,fritscheEB,flambaum25}. To the best of our knowledge,  EB in doped crystals has been so far only investigated in  \ch{Th:CaF_2} \cite{nickerson2020nuclear, bsnpra}.

In this work, we extend the theoretical study of EB to other VUV-transparent crystals recently highlighted in the literature. We investigate how the atomic arrangement surrounding a thorium defect in LiSAF and the chemically similar LiCAF hosts affects the nuclear excitation and deexcitation rates via EB mechanisms. The structural complexity of these materials leads to multiple low-energy defect structures, in some cases influenced by the chemical environment during crystal growth, and requires larger supercells to accurately capture the charge compensation schemes in the dilute doping limit, leading to an increase in both computational demand and simulation accuracy compared to previous studies on  \ch{Th:CaF_2}.

We examine both spontaneous and laser-assisted EB schemes for nuclear excitation and deexcitation for several charge compensation mechanisms. The calculated spontaneous rates are 
orders of magnitude smaller than those for radiative nuclear decay or internal conversion via defect states \cite{morgan2024theory}. However, our results for laser-assisted scenarios show that certain charge compensation schemes in Th:LiCAF and Th:LiSAF provide a set of electronic states that enable nuclear excitation rates more than two orders of magnitude higher than those achieved through direct nuclear excitation with a VUV laser.
Moreover, these charge compensation mechanisms facilitate a quenching scheme that depopulates the isomeric state significantly faster than the radiative decay. This turns out to be beneficial to decrease the duration of interrogation cycles in a nuclear clock.

Our findings are discussed in the context of recent experimental observations in $^{229}$Th:LiSAF \cite{elwell2024laser, terhune2024photo} and $^{229}$Th:CaF$_2$ \cite{schaden2024laser}. 
There are indications that laser-assisted quenching via EB may have occurred in a recent experiment \cite{schaden2024laser}. To check this possibility, we propose an excitation protocol that can confirm the occurrence of EB.

This paper is structured as follows. Section~\ref{sec:Theory} provides a detailed introduction to EB and laser-assisted EB schemes. The characteristics of our model systems, $^{229}$Th:LiCAF and $^{229}$Th:LiSAF, are then presented and analyzed in Sec.~\ref{sec:Crystal}.
Section \ref{sec:results} presents and discusses our numerical results for both spontaneous and laser-assisted EB rates. The results are discussed in the context of current experimental findings in Sec.~\ref{sec:experiment}.
Finally, the paper concludes with a brief discussion in Sec.~\ref{sec:end}.
Atomic units ($\hbar = m_e = e = 1$) are used throughout the present paper unless stated otherwise.

%%%%%%%%%%%%%%%%%%%%%%%%%%%%%%%%%%%%%%%%%%%%%%%%%%%%%%%%%%%%

\section{EB in large band-gap crystals}
\label{sec:Theory}

This Section gives a brief overview of EB schemes in large band-gap crystals for both nuclear excitation and deexcitation.
We start by introducing the spontaneous EB schemes and their theoretical description. Here, we adapt the formalism presented in Refs.~\cite{nickerson2020nuclear,bsnpra} and introduce improvements to better take into account the periodic nature of the crystal environment. We then proceed to address laser-assisted schemes. Dimensionless coefficients are introduced and later used to estimate the strength of nuclear (de)excitation via EB compared to other mechanisms.

\subsection{Spontaneous EB schemes}
\begin{figure}
    \centering
    \includegraphics[width=1\linewidth]{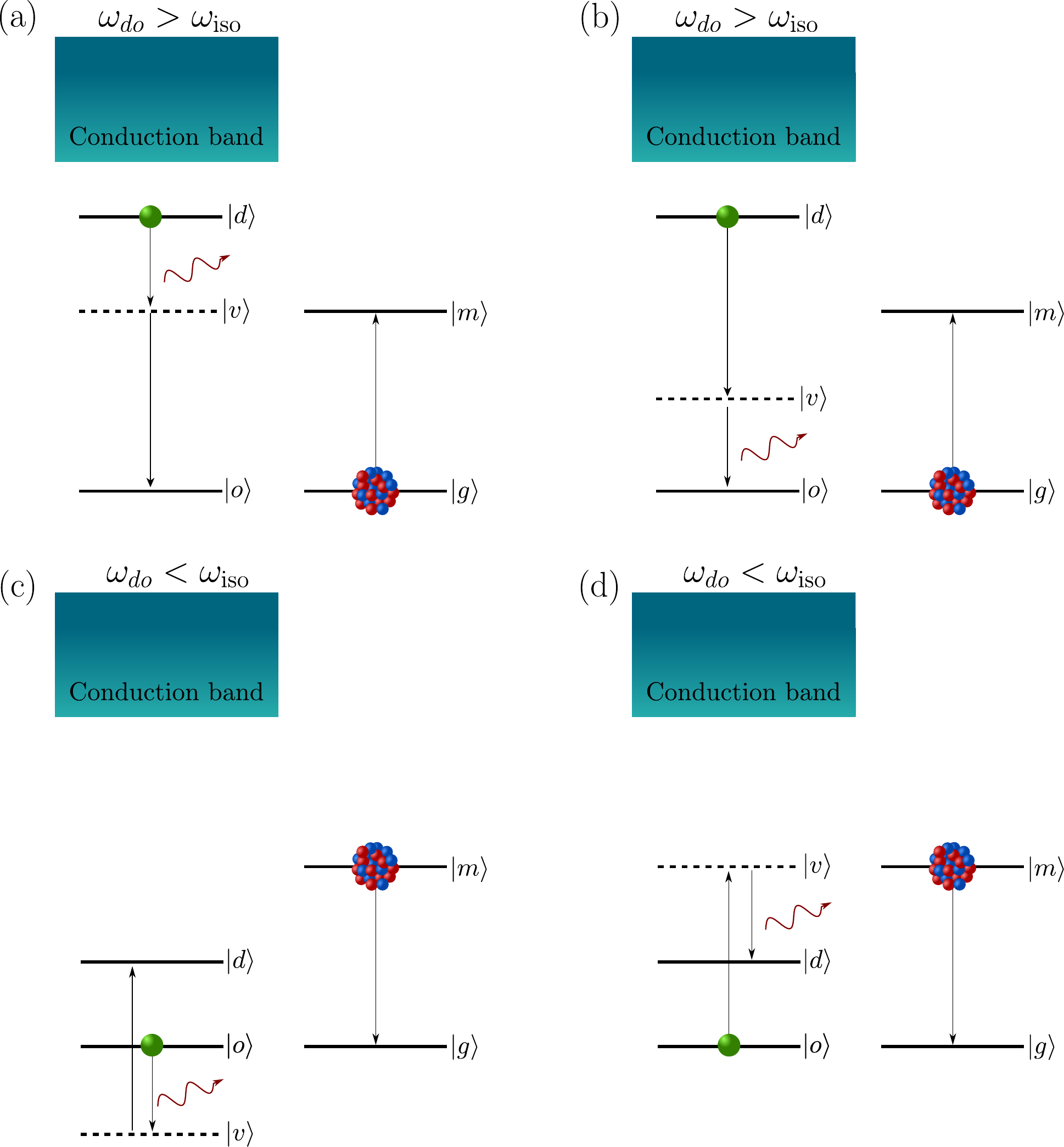}
    \caption{Spontaneous EB schemes: (a), (b) isomer excitation and (c), (d) isomer decay. The electronic shell thereby undergoes transitions between the electronic ground state $\ket{o}$, the virtual state $\ket{v}$, and the defect state $\ket{d}$. $\ket{g}$ ($\ket{m}$) corresponds to the nuclear ground (isomeric) state. The energy of the defect state $\omega_{do}=\omega_d-\omega_o$ relative to the ground state $\ket{o}$ is (a), (b) larger or (c), (d) smaller than the energy of the isomeric state $\omega_{\rm iso}=\omega_m - \omega_g$. }
    \label{spontischemes}
\end{figure}

%%%%%%%%%%%%%%%%%%%%%%%%%%%%%%%%%%%%%
EB refers to a process in which the nucleus is (de)excited via coupling to the electronic shell. 
Figure~\ref{spontischemes} illustrates the underlying mechanism within the crystal environment.
Figure \ref{spontischemes}~(a) depicts a possible scheme for spontaneous excitation via EB. In this scenario, a defect state $\ket{d}$ located above the nuclear isomer energy is initially occupied. The electron decays to a virtual state $\ket{v}$, emitting a real photon, followed by the further decay to the electronic ground state $\ket{o}$, with the simultaneous excitation of the nucleus from its initial ground state $\ket{g}$ to the excited state $\ket{m}$. Alternatively,  the photon emission and nuclear excitation can also occur in reversed order via a different virtual state, as illustrated in Fig.~\ref{spontischemes}~(b).
%, not depicted in the Figure, but accounted for in the calculation.  

%
If the excitation energy of $\ket{d}$ is lower than that of $\ket{m}$, as shown in Figs.~\ref{spontischemes} (c) and (d), spontaneous excitation via EB is energetically forbidden. However, spontaneous decay of the isomer via EB is possible, again via a virtual electronic state. The energy thereof is given either by  the nuclear isomer energy  $\omega_v-\omega_o=\omega_{\rm iso}$, as illustrated in Fig.~\ref{spontischemes} (d),
or the virtual state can lie below the edge of the valence band, with $\omega_d- \omega_v =\omega_{\rm iso}$, as shown in Fig.~\ref{spontischemes}~(c).

This multi-step process can be described as an effective dipole ($E1$) transition. The rate for spontaneous nuclear excitation can be calculated via \cite{scully1997quantum} 
\begin{equation}
\label{eq:spontirate}
\begin{aligned}
        \Gamma_{\text{eb}}\lrb{\ket{g,d}\rightarrow \ket{m,o}} &= \frac{4}{3} \lrb{\frac{\omega_p}{c}}^3 \frac{1}{N_g N_d} \\&\times \sum_{\substack{m,g,\\ o,d}} \ab{\bra{m,o} \Tilde{\bm{\mathcal{Q}}}_{E1}\ket{g,d}}^2
\end{aligned}
\end{equation}
where $\omega_p = \omega_{do}-\omega_{\mathrm{iso}}$ 
is the energy difference between nuclear and electronic levels and hence the energy of the emitted photon,  and $c$ is the speed of light.
The sums over $d$ and $o$ run over the spin degenerate sublevels of each respective electronic state, and the sums over $m$ and $g$ run over the nuclear substates, respectively. Furthermore, 
$N_d$ is the number of spin-degenerate electronic states involved in the process, while $N_g = 2I_g + 1$ is the number of initial nuclear ground states.
The total rate is then the sum over all final states $\ket{o}$.

The bridge operator $\Tilde{\bm{\mathcal{Q}}}_{E1}$ is a spherical tensor operator with spherical components $\{0,\pm1\}$. The matrix elements of this operator are products of three terms corresponding to the three steps of the process \cite{porsev2010effect,nickerson2020nuclear},
\begin{equation}
\label{eq:bridge_matrix_element}
\begin{aligned}
        \bra{m,o} \Tilde{\bm{\mathcal{Q}}}_{E1} \ket{g,d}&= \sum_{\lambda K, q} (-1)^q \Big[\sum_k \frac{\bra{o}\bm{\mathcal{Q}}_{E1}\ket{k}\bra{k}\mathcal{T}_{\lambda K, q}\ket{d}}{\omega_{dk}- \omega_{\text{iso}}} \\
        &+  \frac{\bra{o}\mathcal{T}_{\lambda K,q}\ket{k}\bra{k}\bm{\mathcal{Q}}_{E1}\ket{d}}{\omega_{ok}+\omega_{\text{iso}}}\Big]\\
        &\times \bra{m}\mathcal{M}_{\lambda K, -q}\ket{g}.
\end{aligned}
\end{equation}
Here, the matrix element $\bra{o}\bm{\mathcal{Q}}_{E1}\ket{k}$ describes the photon emission, while $\bra{k}\mathcal{T}_{\lambda K, q}\ket{d}$ stands for the electronic transition which occurs simultaneously with the nuclear transition described by the matrix element $\bra{m}\mathcal{M}_{\lambda K, -q}\ket{g}$. The indices
$\lambda K$ denote the multipolarity of the hyperfine operators $\mathcal{T}_{\lambda K, q}$ and the nuclear transition operators $\mathcal{M}_{\lambda K, -q}$, respectively. Furthermore, the indices $q \in \lrcb{-L, \dots, L}$ correspond to the spherical components of these operators. To account for the virtual states illustrated in Fig.~\ref{spontischemes}, the summations run over all unoccupied intermediate electronic states $\ket{k}$, including the set of conduction band states $\ket{c}$. The valence band is considered to be fully occupied; holes therein should decay much faster than the time scale of the EB process \cite{elwell2024laser}.
The electronic states $\lrcb{\ket{o},\ket{d},\ket{n}}$ correspond to the crystal wave functions for the valence, defect, and conduction band states obtained in a single-particle picture from DFT simulations, as discussed in more details in Sec.~\ref{sec:Crystal}. Please note that we have neglected in the denominators of Eq.~\eqref{eq:bridge_matrix_element} the imaginary part proportional to the respective intermediate state width.

The  $^{229}$Th isomeric transition proceeds via $M1$ + $E2$ multipole mixing \cite{kirschbaum2024photoexcitation}. Consequently, the summation in Eq.~\eqref{eq:bridge_matrix_element} is restricted to these two multipolarities. We note that the multipolarity of the nuclear transition is not to be confused with the $E1$ multipolarity of the emitted or absorbed bridge photon. In principle, 
the bridge photon could also have other multipolarities. However, as shown in Ref.~\cite{bsnpra}, the corresponding EB rates are orders of magnitude suppressed compared to the $E1$ multipolarity, such that we do not consider them in this work.

The electric dipole operator $\bm{\mathcal{Q}}_{E1}$ describes the dipole emission associated with the electronic transition $\ket{d}\rightarrow \ket{v}$. In crystal lattices, periodic boundary conditions require expressing the operator in reciprocal space as $\bm{\mathcal{Q}}_{E1} = -i\nabla_{\bm{k}}$, since its real-space representation is invalid due to the lack of a well-defined origin~\cite{dipoleoperator}. 
In contrast, the hyperfine operators are inherently local due to their $1/r^3$ dependence, as shown below.

For $\lambda K = M1$, the interaction between the nucleus and the electronic shell corresponds to a coupling between the nuclear and electronic charge currents. 
The corresponding non-relativistic coupling operator reads~\cite{abragam1961principles,bransden2006physics}
\begin{equation}
\label{eq:current}
    \mathcal{T}_{M1, q} = \frac{1}{c^2} \lrsb{\frac{L_q}{r^3}- \frac{\sigma_q}{2r^3}+ 3 \frac{r_q(\bm{r}\cdot \bm{\sigma})}{2r^5}+ \frac{4 \pi}{3}  \sigma_q \delta(\bm{r})}%\, ,
\end{equation}
where $L_q$ denotes the orbital angular momentum of the electron, $\bm{\sigma}$ $(\sigma_q)$ are the Pauli matrices and $\delta(\bm{r})$ is the Dirac delta function. 

For $\lambda K = E2$, the nucleus-electron coupling is described by the  Coulomb interaction  \cite{johnson2007atomic}
\begin{equation}
\label{eq:coulomb}
     \mathcal{T}_{E2, q} = -\frac{1}{r^3} \sqrt{\frac{4\pi}{5}} Y_{2,q}(\theta, \phi) 
\end{equation}
with $Y_{2,q}(\theta, \phi)$ being the spherical harmonics.
The matrix element of the nuclear transition operator can be written in terms of its reduced matrix element via \cite{edmonds1996angular}
\begin{equation}
\small
\begin{aligned}
        \bra{I_m \, m_m} \mathcal{M}_{\lambda K, -q} \ket{I_g \, m_g} &=\frac{(-1)^{I_g-m_g}}{\sqrt{2K+1}} \braket{I_g m_g I_m -m_m |K -q} \\
        &\times \langle I_m || \mathcal{M}_{\lambda K} || I_g \rangle
\end{aligned}
\end{equation}
where $\braket{I_g m_g I_m -m_m |K -q}$ is the Clebsch Gordan coefficient, $I$  the the nuclear angular momentum and $m$ its projection, respectively. As a shortened notation, nuclear states are denoted throughout this work also as $\ket{\zeta}\equiv \ket{I_\zeta \, m_\zeta}$ where $\zeta \in \lrcb{g,m}$.
The reduced matrix element is related to the reduced transition probability $B(\lambda K, I_g \rightarrow I_m)\equiv B_\uparrow(\lambda K)$ according to \cite{porsev2010effect}
\begin{equation}
    |\langle I_m || \mathcal{M}_{\lambda K} || I_g \rangle| = \sqrt{\frac{4\pi(2I_g +1)}{2K+1}B_\uparrow(\lambda K)}.
\end{equation}
Note that the current-current operator in Ref.~\cite{bsnpra} has a $1/c$ dependence while the operator presented here in Eq.~\eqref{eq:current} has a $1/c^2$ dependence. This difference is related to the definition of reduced transition probability of a magnetic dipole transition, which is proportional to the nuclear magneton $\mu_N$. In Ref.~\cite{bsnpra}, the latter is given by $\mu_N = e \hbar/(2m_p c)$ while in the present work the definition $\mu_N = e \hbar/(2m_p)$ is used.
For the reduced transition probabilities, we use the values for the decay $B_\downarrow (M1) =0.022 \, \text{W.u.}$ reported in \cite{schaden2024laser,fritscheEB} and $B_\downarrow (E2)= 27.04 \, \text{W.u.}$ predicted in \cite{minkov2017reduced}.

\subsection{Laser assisted schemes}
\begin{figure}
    \centering
    \includegraphics[width=1\linewidth]{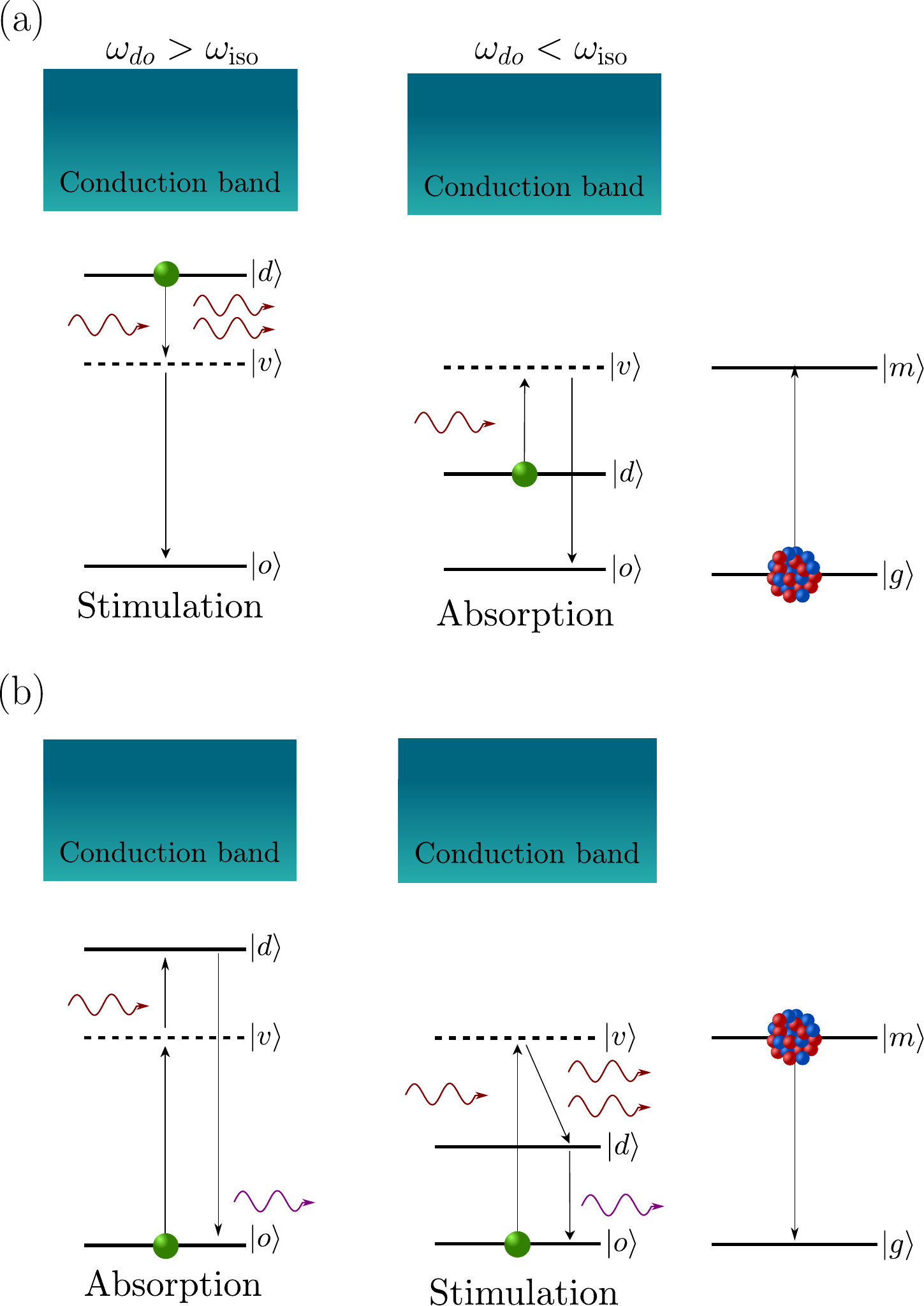}
    \caption{Laser-assisted EB schemes. (a) Driven EB schemes for nuclear excitation. Depending on the energy of the defect state, one can distinguish between stimulation, where an external laser photon facilitates the excitation via stimulated emission, and absorption, where an external laser photon renders the EB process possible. (b) Quenching schemes for nuclear deexcitation.}
    \label{fig:laser_assited}
\end{figure}

%%%%%%%%%%%%%%%%%%%%%%%%%%%%%%%%%%%%%%%%%%%%%%%%
The EB process for nuclear (de)excitation can also be assisted by an optical$/$UV laser. We refer to laser-assisted nuclear excitation as driven EB schemes, and laser-assisted nuclear deexcitation as EB quenching. The processes that we investigate all consider the case $\omega_v - \omega_o = \omega_{\mathrm{iso}}$  and are schematically illustrated in Fig.~\ref{fig:laser_assited}.
Here, one can distinguish between stimulated emission and absorption due to the laser photons, depending on whether the energy of the defect state relative to $\ket{o}$ is larger or smaller than that of the nuclear isomer, and on the initial nuclear state.
We assume in the following that a continuous wave (cw) laser is assisting the EB process with photons at the same frequency as the EB photon.
The rate of a laser-assisted stimulation process of the electronic transition  $\ket{d}\rightarrow \ket{v}$  can be related to the rate of the corresponding spontaneous process via (in SI units) \cite{sobelman2012atomic}
\begin{equation}
\label{eq:stim}
  \Gamma_{\text{st}}(\ket{g,d}\rightarrow\ket{m,o}) = \Gamma_{\text{sp}}(\ket{g,d}\rightarrow\ket{m,o}) \frac{\pi^2 c^2 }{\hbar \omega_p^3} I_\omega 
\end{equation}
where $I_\omega$ is the spectral intensity of the laser, which is given by
\begin{equation}
    I_\omega = \frac{\mathcal{I}}{\Gamma_\ell}\,.
\end{equation}
Here, $\mathcal{I}$ is the intensity of the laser, assumed not to be a true monochromatic source, in its continuous wave (cw) equivalent, and $\Gamma_\ell$ is the corresponding non-zero line width, respectively. 
The stimulated rate $\Gamma_{\text{st}}$  can be related to the  absorption process rate for the electronic transition $\ket{v}\rightarrow \ket{d}$ (provided the defect state lies for this case below the isomer energy) as 
\begin{equation}
    \Gamma_{\text{ab}}(\ket{g,d}\rightarrow\ket{m,o})=   \frac{N_m N_o}{N_g N_d}\Gamma_{\text{st}}(\ket{m,o}\rightarrow\ket{g,d})
\end{equation}
where $N_o/N_d$ denotes the ratio of degeneracies of the involved electronic states and $N_m/N_g$ of the nuclear states, respectively. Thus, as an input, we must first calculate the spontaneous EB process of interest.
In order to evaluate the (dis)advantage compared to direct laser driving of the nuclear transition, we define the dimensionless coefficient
\begin{equation}
\label{eq:enhance}
    \beta_\ell = \frac{\rho_d \Gamma_D (\ket{g,d}\rightarrow\ket{m,o})}{\Gamma_{\text{ex}}}
\end{equation}
where $\rho_d$ is the steady-state occupation probability of the defect state involved in the EB process and the index $D=\{\rm st, ab\}$ denotes the type of laser driven scheme (stimulation or absorption). In the denominator,  $\Gamma_{\text{ex}}$ denotes the direct nuclear photo-excitation rate. 

In a similar fashion, we can define the quenching coefficient via
\begin{equation}
\label{eq:quench}
    \alpha_{\mathrm{qu}} = \frac{\Gamma_Q(\ket{m,o}\rightarrow\ket{g,d})}{n^3 \Gamma_\gamma}
\end{equation}
where $Q\in \lrcb{\mathrm{st}, \mathrm{ab}}$ is the type of the quenching scheme and $\Gamma_\gamma$ is the radiative decay rate of the bare nucleus and $n$ the index of refraction of the medium \cite{tkalya2000spontaneous,tkalya2000decay}.

\section{Electronic structure of the crystal environment}
\label{sec:Crystal}
%%%%%%%%%%%%%%%%%%%%%%%%%%%%%%%%%%%%%%%%%%%%%%%%%%%%%%%%%%%%%%%%%

The calculation of EB rates in a crystalline environment relies on using crystal wave functions, which describe the electronic surroundings of the thorium nucleus. In this Section, we introduce our methodology to calculate these wave functions and provide a summary of the results presented in Ref.~\cite{martin}, which address the electronic structure of $^{229}$Th:LiCAF and $^{229}$Th:LiSAF.

Our simulations employ the Vienna Ab-initio Simulation Package (VASP)~\cite{vasp1, vasp2, vasp3, vasp4, vasp5} in a plane wave basis and the projector augmented wave method \cite{paw}. We evaluate the Kohn-Sham wavefunctions in the scalar relativistic approximation~\cite{ATechniqueForKoelli1977, TheScalarRelaTakeda1978}, and treat the electrons' spins as collinear. Hence, the four-component wave function reduces to just two components, one for each spin direction.

We calculate these single-particle wave functions by first obtaining the plane wave expansion coefficients $C_{n\bm{k}\bm{G}}$ from the converged VASP simulation,  where $n$ is the band index, $\bm{k}$ is a vector in the first Brillouin zone, and $\bm{G}$ is a reciprocal lattice vector. Then, we calculate pseudo-wave functions on a spherical grid
\begin{equation}
\tilde{\psi}_{n\bm{k}}(\bm{r}) = \Omega^{-1/2} \sum_G C_{n\bm{k}\bm{G}} e^{-i(\bm{k} + \bm{G}) \cdot \bm{r}},     
\end{equation}
where $\Omega$ is the volume of the Wigner-Seitz cell~\cite{LowEnergyElecFeenst2013, PhysRev.43.804}.  The density of radial grid points increases exponentially with distance to the origin, while we use angular points on regular intervals. From the pseudopotential VASP input file (\texttt{POTCAR}) for thorium, we read the radial components of the all-electron and pseudo partial waves, $u_i$ and $\tilde{u}_i$, and for each structure we extract projectors $\braket{p_i | \tilde{\psi}_{n\bm{k}}}$ to reconstruct the normalized all-electron Kohn-Sham wave function
\begin{equation}
    \psi_{n\bm{k}}(\bm{r}) = \tilde{\psi}_{n\bm{k}}(\bm{r}) + \sum_i [u_i(r) - \tilde{u}_i(r)]\braket{p_{i\bm{k}} | \tilde{\psi}_{n\bm{k}}} Y_{l, m}(\theta, \phi)/r,
\end{equation}
where $i$ is a composite index of quantum numbers $n$, $l$, $m$. 
%\tk{Is $Y_{l, m}(\theta, \phi)/r$ or $Y_{l, m}(\theta, \phi)$ correct? } \pamp{$Y_{l,m}/r$ is correct.}
In our calculations, we use the Perdew–Burke–Ernzerhof (PBE) \cite{pbe} exchange correlation potential. We select a kinetic energy cutoff in our plane wave basis set of 950 eV and evaluate the doped supercells at $\bm{k} = \Gamma$. Our supercells are $4 \times 4 \times 2$ repetitions of the unit cell of LiCAF or LiSAF, resulting in a cell with lattice vectors of approximately equal length and around 576 atoms.

We calculated defect formation energies for temperatures ranging from 0 K to the melting point of LiCAF and LiSAF, while controlling the fluorine availability via its chemical potential from saturation to deficiency \cite{martin}. The two materials exhibit distinct behaviors. When describing defect formation energies, we use Kröger-Vink notation~\cite{RelationsBetweKroger1956}   and a condensed notation composed of five numbers corresponding to the species Li, Ca (Sr), Al, F, and Th. A number with a bar indicates the removal of an atom from the supercell, while a number without a bar denotes its addition.

The chemical potential of fluorine influences the energetically preferred charge compensation scheme of \ch{Th:LiCAF}. Given that growth conditions frequently involve a fluorinated environment, we will focus on that scenario. Among the identified structures, $\ch{Th_{Ca}^{..}} + \ch{2 v_{Li}^'}$ ($\bar{2}\bar{1}001$) clearly exhibits the lowest energy. The next three lowest compensation schemes are $\ch{Th_{Ca}^{..}} + \ch{v_{Ca}^{''}}$ ($0\bar{2}001$), $\ch{Th_{Ca}^{..}} + \ch{Li_{Al}^{''}}$ ($1\bar{1}\bar{1}01$), and $\ch{Th_{Al}^{.}} + \ch{v_{Li}^{'}}$ ($\bar{1}0\bar{1}01$). Notably, the latter two schemes are also preferred in fluorine-deficient environments.

Contrary to \ch{Th:LiCAF}, the fluorine environment has a negligible influence on the lowest energy charge compensation scheme of \ch{Th:LiSAF}, which is $\ch{Th_{Al}^{.}} + \ch{v_{Li}^{'}}$ ($\bar{1}0\bar{1}01$). The other low energy compensation schemes investigated in this study are $\ch{Th_{Sr}^{..}} + \ch{Li_{Al}^{''}}$ ($1\bar{1}\bar{1}01$), $\ch{Th_{Al}^{.}} + \ch{F_i^{'}}$ ($00\bar{1}11$), and $\ch{Th_{Sr}^{..}} + \ch{2 F_i^{'}}$ ($0\bar{1}021$).

In the following, we calculate electronic bridge rates for these selected low-energy structures
in Th:LiCAF and Th:LiSAF. A key objective is to identify whether these rates exhibit distinct behavior. If so, structures with a higher EB rate may be exploited by adjusting element concentrations during crystal growth to selectively favor specific compensation mechanisms. Furthermore, depending on the precise energy of defect states in a given structure, a targeted bridge (de)excitation may be achieved through selective laser irradiation in driven EB schemes.

\section{Numerical results}
\label{sec:results}
%%%%%%%%%%%%%%%%%%%%%%%%%%%%%%%%%%%%%%%%%%%%

We calculate the rates of both spontaneous and laser-assisted EB schemes for four of the lowest-energy charge compensation schemes of $^{229}$Th:LiCAF and $^{229}$Th:LiSAF.  
 For each charge compensation scheme, we calculate the DFT energies and wave functions for several thousand unoccupied states up to 25 eV above the Fermi energy.  The Fermi level is chosen as the electronic ground state, since a hole in the valence band is expected to relax to the Fermi level on a much shorter time scale than the bridge process \cite{elwell2024laser}. Detailed checks presented in 
Appendix~\ref{sec:appendix1} show that approx.~2000 unoccupied states are sufficient in the summation over intermediate states to reach convergence of the EB rate.

The employed crystal wave functions are eigenstates of neither angular momentum nor parity operators; instead, the one-electron wave functions are defined by their energy and $\bm{k}$-point. These wave functions are represented on a spherical grid centered around the thorium nucleus. The grid consists of $N = N_r \cdot (N_\theta - 2) \cdot N_\phi + 2N_r$ points, with unique points in each direction given by $(N_r, N_\theta, N_\phi)=(349,26,50)$. A constant spacing is assumed for the angular components, while the radial spacing follows $r_n = r_0 e^{n/\kappa}$, where $r_0 = \num{1.35e-4} a_0$ ($a_0$ being the Bohr radius) and $\kappa=31.25$. Larger grids, such as $(N_r, N_\theta, N_\phi)=(349,34,66)$, were also tested but did not yield any significant improvement in the accuracy of the results.

Because the spherical grid samples a subset of the entire supercell, the normalization $\braket{\psi_n | \psi_n} = 1$ is violated for delocalized wave functions. While the normalization could be achieved using a Cartesian grid, such an approach results in poor sampling near the thorium position. Given that the hyperfine operators exhibit a $1/r^3$ dependence, fine sampling in this region is crucial, and is provided by our choice of radial grid. Note that the dipole moments $\bra{\psi_{n\bm{k}}}-i\nabla_{\bm{k}}\ket{\psi_{n'\bm{k}}}$ are calculated directly in VASP in the reciprocal space and therefore remain unaffected by the spatial normalization issue~\cite{Gajdos2006}. 

According to the calculated norm in the thorium-centered spherical grid, we can classify the defect states as localized or delocalized around the thorium nucleus.
Here, we designate states with $\braket{\psi_n | \psi_n} \ge 0.4$ as thorium-localized and $\braket{\psi_n | \psi_n} < 0.4$ as thorium-delocalized. We further regard each of these states, up to the localized defect state with the highest energy, as the initial (or final) state in our EB rates calculation (see Figs.~\ref{fig:sponti} and~\ref{fig:laassist}).

Unfortunately, predicted DFT one-electron energies are known to have limited accuracy, failing to reproduce experimentally measured band gaps~\cite{DensityFunctioPerdew2009}. We therefore do not expect them to provide reliable estimates of whether defect states lie above or below the nuclear transition energy. To mitigate this systematic error, we apply two strategies. 
In Sec.~\ref{sec:sponti} and \ref{sec:la} a rigid shift $\Delta_r$ is applied to the unoccupied states' energies, ensuring that the experimentally determined band gap energies $E_g$ for pristine LiCAF and LiSAF are reproduced, leading to $\Delta_r = E_g^{\text{exp}} - E_g^{\text{DFT}}$. Note that reliable data on the band gap of doped crystals is not available at present. Thus, for the scope of this work, we use the pristine crystal band gap values 11.07~eV for LiCAF \cite{shimamura2000growth}, and 10.69~eV for LiSAF~\cite{sakai1999licaf}.
As a second strategy, in addition to the rigidly shifted unoccupied states, in Sec.~\ref{sec:vary} we apply a further variable energy shift to unoccupied electronic states whose energies encompass the range from the lowest to the highest unoccupied thorium-localized state directly following the band gap.  By distinguishing between localized and delocalized states through this approach, we acknowledge that DFT errors may differ significantly across these categories~\cite{AbInitioStudyPimon}; however, by including some delocalized states with variable energy, we also consider their potentially larger dipole transition matrix elements' contribution to the electronic bridge rate (to be discussed in Sec.~\ref{sec:sponti} and Appendix~\ref{appendix}).

\subsection{Spontaneous schemes}
\label{sec:sponti}
\begin{figure*}
    \centering
\includegraphics[width=1.02\linewidth]{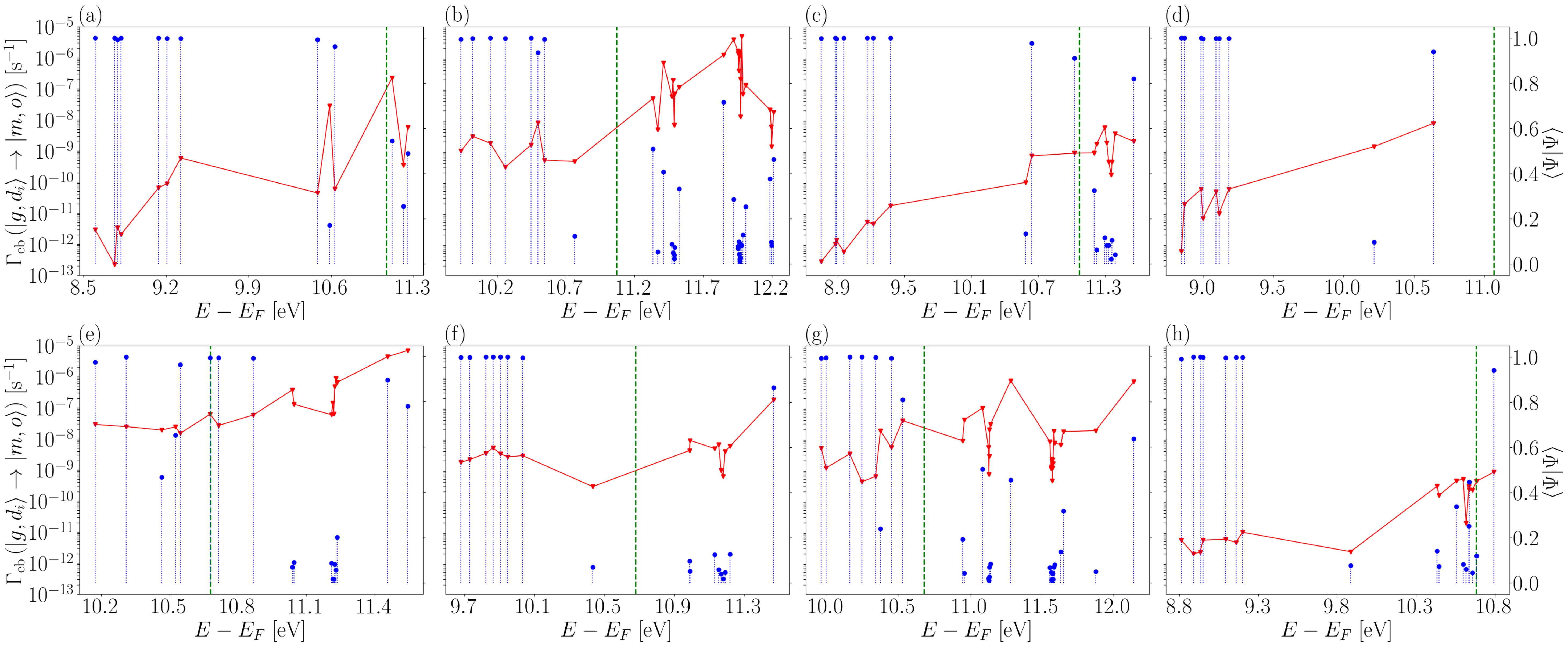}
    \caption{Spontaneous EB nuclear excitation rates $\Gamma_{\rm eb}(\ket{g,d_i}\rightarrow\ket{m,o})$(red, left-hand scale) and degree of localization (blue, right-hand scale) as a function of initial defect state energy relative to the Fermi level $E_F$. The upper row corresponds to $^{229}$Th:LiCAF with compensation schemes  (a) $0\bar{2}001$, (b) $\bar{1}0\bar{1}01$, (c) $\bar{2}\bar{1}001$, (d) $1\bar{1}\bar{1}01$. The lower row corresponds to $^{229}$Th:LiSAF with compensation schemes (e) $00\bar{1}11$, (f) $0\bar{1}021$, (g) $\bar{1}0\bar{1}01$, (h) $1\bar{1}\bar{1}01$.}
    \label{fig:sponti}
\end{figure*}

%%%%%%%%%%%%%%%%%%%%%%%%%%%%%%%%%%%%%%%%%%%%
Figure~\ref{fig:sponti} presents the spontaneous EB rates for nuclear excitation ($\omega_{do}> \omega_{\mathrm{iso}}$) for the four lowest-energy configurations of both Th:LiCAF [upper row, panels (a)-(d)] and Th:LiSAF [lower row, panels (e)-(h)].
 Each red bullet point corresponds to EB rates calculated for a different initial defect electronic state $d_i$.  The rates are plotted against the defect energy relative to the Fermi level, $E_{d_i}-E_F$. The blue bullets indicate the norm of the respective state. The green dashed line marks the conduction band edge, which separates the band gap from the conduction band. Through the applied rigid energy shift $\Delta_r$, the conduction band edge energies are fixed at 11.07~eV for LiCAF, and 10.69~eV for LiSAF.

Each charge compensation scheme exhibits a different number of localized and delocalized electronic states in and slightly above the band gap, leading to different total values of the EB rate. 
According to our designated procedure, we consider as initial state $d_i$ also states falling within the conduction band (after the rigid shift), up to the localized state with highest energy.  This appears justified since neither the precise energies nor a reliable estimate of the band gap are available.

The spontaneous EB rates range from $\mathcal{O}(\num{e-13}\si{\per \second})$ to $\mathcal{O}(\num{e-6}\si{\per \second})$. A general trend observed across most structures is that larger EB rates emerge at higher initial state energies ($E - E_F > \Delta_g$), particularly for localized defect states. This is the result of several factors. Consider, for instance, the LiCAF structure $0\bar{2}001$ [Fig.~\ref{fig:sponti} (a)]. The smallest EB rate is obtained when considering as initial state a defect state with $E_d = \SI{8.76}{\electronvolt}$ where $\Gamma_{\mathrm{eb}}^{\mathrm{min}} \sim \mathcal{O}(\SI{e-13}{\per \second})$. While this state has strong hyperfine coupling matrix elements (see Appendix \ref{appendix}), the corresponding photon energy is only $\omega_p \sim \SI{0.4}{\electronvolt}$.

In contrast, the highest EB rate in the system occurs for a defect state at $11.12$ eV, where $\Gamma^{\mathrm{max}}_{\mathrm{eb}} \sim \SI{2e-7}{\per \second}$. Its norm is $\braket{\Psi | \Psi} \sim 0.56$, and its hyperfine coupling is on the same order of magnitude as that of fully localized states (see Appendix \ref{appendix}). However, in this case, a photon with 2.76~eV energy is created. The ratio $\left(\omega^{\mathrm{max}}_{p}/\omega^{\mathrm{min}}_{p}\right)^3 \sim \num{3e2}$ already leads to an increase of more than two orders of magnitude. Furthermore, several terms in the intermediate summation span a few orders of magnitude as shown in Fig.~\ref{fig:dipolematrix} in  Appendix \ref{appendix}. In combination with the range of energy differences entering the denominator in Eq.~\eqref{eq:bridge_matrix_element}, the large spread of the EB rates appears viable and is not the result of a numerical artifact.

Large EB rates for strongly delocalized states ($\braket{\Psi | \Psi} \lesssim 0.1$), as seen in Fig.~\ref{fig:sponti} (b), arise primarily due to large dipole moments that compensate for the weak hyperfine coupling. 
The large dipole moments stem from the dispersion of the involved electronic state in the reciprocal space.
In most cases, delocalized states exhibit a steep dispersion, whereas localized defect states are characterized by a flat dispersion.
Since we compute dipole moments using a derivative in momentum space, a steeper dispersion naturally results in larger dipole moments.

From these observations, we conclude that the degree of localization as reflected in the value of the hyperfine coupling matrix element plays a secondary role in the magnitude of an EB rate. Instead, dipole moments and the energetic positioning of states are the dominant factors. This implies that even delocalized states can contribute significantly to nuclear (de)excitation via EB in a crystal.

Regarding the crystal type, $^{229}$Th:LiSAF offers a more favorable environment for observing EB, as three out of four charge compensation schemes present larger spontaneous EB rates compared to $^{229}$Th:LiCAF.
However, the absolute magnitude of the spontaneous EB rate remains very small compared to radiative decay or internal conversion \cite{morgan2024theory, thirolfIc25}. Consequently, for future clock operation or interrogation protocols, spontaneous EB is expected to play only a minor role.

\subsection{Laser assisted schemes}
\label{sec:la}
%%%%%%%%%%%%%%%%%%%%%%%%%%%%%%%%%%%%%%
The potential of laser-assisted EB schemes for clock operation and interrogation protocols has been discussed in the literature so far for \ch{Th:CaF_2} \cite{nickerson2020nuclear,bsnpra,schaden2024laser}. 
Theoretically, they should facilitate significantly stronger excitation rates compared to direct laser excitation of the nucleus or much faster nuclear depopulation. In this Subsection, 
we calculate rates of driven EB schemes in LiCAF and LiSAF considering the stimulation scheme for nuclear excitation and the absorption scheme for quenching as illustrated in Fig.~\ref{fig:laser_assited}, both requiring an initial (respectively final) defect state above the isomer energy. We base our analysis on the predicted DFT energies with applied rigid shift $\Delta_r$. 

We start with the case of the stimulation scheme for nuclear excitation. 
 To determine the dimensionless enhancement coefficient $\beta_{\ell}$ in Eq.~\eqref{eq:enhance}, we require knowledge of the population of the defect state involved in the bridge process.
The electronic state population involved in the EB process can be calculated via the steady-state occupation probability of a two-level system, which is given by  \cite{von2020theory}
\begin{equation}
    \rho_d = \frac{\Omega^2}{\frac{2 \Gamma}{\Tilde{\Gamma}}\lrb{\Delta^2 + \Tilde{\Gamma}^2} + 2 \Omega^2}.
\end{equation}
Here, $\Omega = \bm{\mathcal{Q}}_{E1} \cdot \bm{E} / \hbar$ (SI units) denotes the Rabi frequency for an $E1$ transition, $\bm{E}$ is the laser electric field strength and $\Gamma$ represents the radiative relaxation rate of the electronic state. The latter can be computed via Eq.~\eqref{eq:spontirate} by replacing $ \bra{m,o} \Tilde{\bm{\mathcal{Q}}}_{E1} \ket{g,d}$ with $ \bra{o} \bm{\mathcal{Q}}_{E1} \ket{d}$ and consequently dropping the summation over the nuclear state sublevels. The total decay rate of the coherences is given by $\Tilde{\Gamma} = (\Gamma + \Gamma_\ell+ 2\Gamma_{\mathrm{dec}})/2$, where $\Gamma_\ell$ denotes again the linewidth of the laser, $\Gamma_{\mathrm{dec}}$  the internal decoherence (e.g. phonon scattering), and $\Delta$  the detuning, respectively. The radiative lifetime of these electronic states varies between a few tens of nanoseconds and several hundred microseconds. In the following, we will assume that $\Tilde{\Gamma} \approx \Gamma_\ell/2$, since this corresponds to the dominant source of decoherence for the current available broad-band VUV lasers.
For the calculation of the Rabi frequency, we assume that the dipole moment and the electric field are aligned, providing (in SI units) \begin{equation} 
\Omega = \sqrt{\frac{6\pi c^2 \Gamma \mathcal{I}}{\hbar \omega^3}} , 
\end{equation} where we express the electric field in terms of the intensity $\mathcal{I}$ of the driving field as $E = \sqrt{2\mathcal{I}/c\varepsilon_0}$, with $\varepsilon_0$  the vacuum permittivity.

The defect energies range from approximately 8.6 eV to 12 eV for both crystals (see Fig.~\ref{fig:sponti}). To excite these defects, we consider the pulsed four-wave mixing VUV laser discussed in Refs.~\cite{thielking2023vacuum,tiedau2024laser,schaden2024laser}. While this laser is tunable only between 7.4 eV and 10.2 eV, we assume, as a crude approximation, that it would maintain similar performance above 12 eV. This ensures that each level is driven with the same laser power. This assumption is reasonable given the uncertainties in the exact energies of these electronic states, which might well lie in the tunability interval. 
The cw equivalent intensity of this laser amounts to $I_{\mathrm{cw}} = \SI{1.4e4}{\watt \per \metre \squared}$, based on a peak pulse energy of \SI{15}{\micro \joule}, a pulse duration of $\SI{10}{\nano \second}$, and a repetition rate of $\SI{30}{\hertz}$, focused to a beam waist of \SI{100}{\micro \metre}. In combination with the linewidth of the laser $\Gamma_\ell = 2\pi\times \SI{10}{\giga \hertz}$, this setup yields nuclear excitation rates of approximately $\Gamma_{\mathrm{ex}} \approx \SI{3e-7}{\per \second}$.

We also calculate EB rates for the quenching process, in which nuclear deexcitation becomes possible through laser photon absorption as depicted in the left-hand side graph of Fig.~\ref{fig:laser_assited} (b). The defect state plays in this case the role of the final state, and no prior excitation thereof is considered.
For the bare radiative lifetime required for the calculation of the quenching coefficient \eqref{eq:quench}, we adopt $\tau = 1800$ s, which corresponds to the mean value of the measurements reported in Refs.~\cite{tiedau2024laser, elwell2024laser}. The refractive indices of both LiCAF and LiSAF can be estimated as $n = 1.49$ \cite{woods1991thermomechanical, jeet2018search,elwell2024laser} at photon energies of $E=\SI{8.36}{\electronvolt}$, leading to a modified radiative decay rate of $n^3\Gamma_\gamma = \SI{1.8e-3}{\per \second}$.

For both laser-assisted schemes, the wavelength of the external laser falls within the infrared to weak UV spectral range. Since powerful coherent sources are readily available in this domain, the spectral intensity is, in principle, only limited by practical considerations. However, a balance must be maintained between power and linewidth to prevent damage to the crystal and to avoid excessively long scanning times, given that the bridge resonance is extremely narrow.
Therefore, we adopt a spectral intensity of $I_\omega= \SI{5}{\watt \second \per \metre \squared}$, which represents a realistic value for this wavelength range. It is important to note that the dimensionless coefficients $\beta_\ell$ and $\alpha_{\rm qu}$ scale linearly with the spectral power, allowing for some flexibility.

\begin{figure*}
    \centering
   \includegraphics[width=1.02\linewidth]{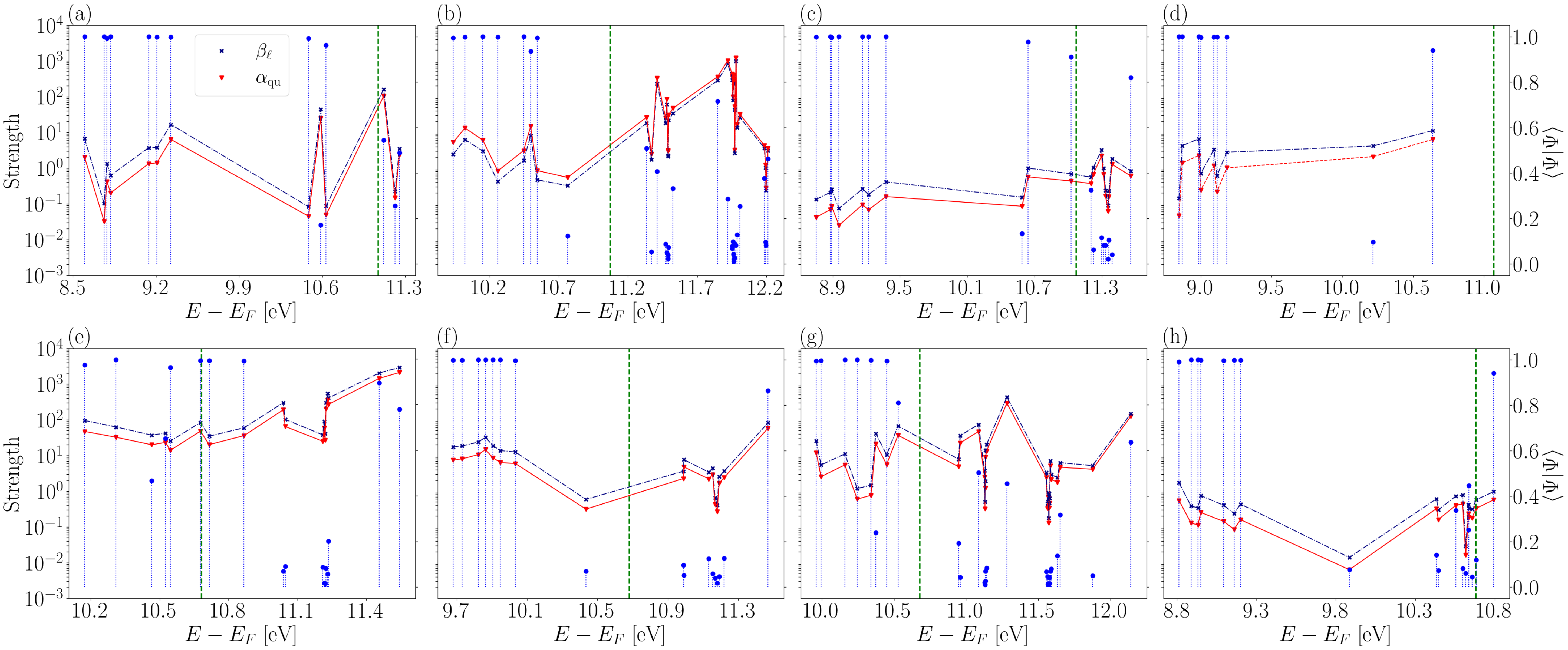}
    \caption{Dimensionless enhancement coefficients $\beta_\ell$, $\alpha_{\mathrm{qu}}$ (left-hand scale) and degree of localization (blue, right-hand scale) as a function of initial/final defect state energy relative to the Fermi level. The upper row corresponds to for the four lowest energy charge compensation schemes in $^{229}$Th:LiCAF (a) $0\bar{2}001$, (b) $\bar{1}0\bar{1}01$, (c) $\bar{2}\bar{1}001$, (d) $1\bar{1}\bar{1}01$.  The lower row corresponds to the $^{229}$Th:LiSAF compensation schemes (e) $00\bar{1}11$, (f) $0\bar{1}021$, (g) $\bar{1}0\bar{1}01$, (h) $1\bar{1}\bar{1}01$. See text for the  considered laser parameters. }
    \label{fig:laassist}
\end{figure*}

We present the calculated dimensionless enhancement coefficients for laser-driven excitation $\beta_\ell$ and quenching $\alpha_{\mathrm{qu}}$ in Figs.~\ref{fig:laassist} and summarize our findings in Tab.~\ref{tab:la}. Similarly to the spontaneous rates results presented in Fig.~\ref{fig:sponti}, we plot the dimensionless coefficients for the considered four charge compensation mechanisms for the two crystals, as a function of the energy of the initial/final defect state. In addition, the degree of localization for the respective initial/final state is presented in the form of the wave function norm $\langle \Psi|\Psi \rangle$. We summarize the span in order of magnitude by listing the maximum and minimum values of the coefficients, along with the corresponding energy and relaxation rate to the Fermi level, in the Table.

For the driven schemes, we use $\Gamma_{\mathrm{ex}}=\SI{3e-7}{\per \second}$, $\mathcal{I}_{\mathrm{cw}}= \SI{1.4}{\watt \per  \metre \squared}$ and $\Gamma_\ell = 2 \pi \times \delta_\nu $ where $\delta_\nu =\SI{10}{\giga \hertz}$. For the quenching schemes, we use $n^3 \Gamma_\gamma =\SI{1.8e-3}{\per \second}$ for both crystals. For both coefficients, $I_\omega=\SI{5}{\watt \second \per \metre \squared}$ is used for the spectral intensity of the assisting laser. In all calculations, resonance was assumed ($\Delta =0$). The steady-state population of the defect is on the order of $\rho_d \sim \mathcal{O}(\num{e-4})$ throughout the energy range of 8.6 eV to 12 eV.

\begin{table*}[htpb!]
    \centering
    \begin{tabular*}{\linewidth}{@{\extracolsep{\fill}} cc|cccc|cccc}
    \hline 
    \hline
  Crystal & Structure &  $\beta_\ell^{\mathrm{min}}$ & $\alpha_{\mathrm{qu}}^{\mathrm{min}}$ & $E-E_F$ [\si{\electronvolt}]& $\Gamma$ [\si{\per \second}] &   $\beta_{\ell}^{\mathrm{max}}$ & $\alpha_{\mathrm{qu}}^{\mathrm{max}}$  & $E-E_F$ [\si{\electronvolt}] & $\Gamma$ [\si{\per \second}]\\ 
  \hline
   $^{229}$Th:LiCAF & $0\bar{2}001$ &  \num{0.10}& \num{3.2e-2} & 8.76 & \num{1.4e4} & \num{1.6e2} & \num{1.0e2} & 11.12& \num{1.1e6} \\
  &$\bar{1}0\bar{1}01$&\num{0.28} & \num{0.24} & 12.20 & \num{1.3e6}  & \num{1.2e3}& \num{9.9e2} & 11.98 &\num{2.4e7}\\
  & $\bar{2}\bar{1}001$& \num{7.4e-2} & \num{2.5e-2} & \num{8.96} & \num{1.3e5} & \num{3.2} & \num{2.2}   &11.30 & \num{5.3e5}\\
  &$1\bar{1}\bar{1}01$ & \num{0.14}& \num{4.7e-2} & \num{8.84} & \num{9.4e3} &  \num{11.19} & \num{6.3}  & 10.64 & \num{1.4e5} \\
  \hline 
     $^{229}$Th:LiSAF& $00\bar{1}11$ & 25.3&13.9 & 10.54  & \num{1.0e7} & \num{2.9e3}  & \num{2.1e3}  & 11.54 & \num{1.0e7} \\
  &  $0\bar{1}021$& \num{0.41}& 0.27  & \num{11.18} & \num{5.7e6}  &  82.3  & 58.3 &11.46 & \num{1.4e6} \\
 &  $\bar{1}0\bar{1}01$& \num{0.18} & 0.13  & 11.57 & \num{2.4e5}& \num{4.3e2}   & \num{2.9e2} & 11.28 &  \num{2.4e6} \\
  & $1\bar{1}\bar{1}01$ &  \num{1.4e-2}& \num{6.3e-3} & 9.88 &  \num{8.4e6}  & 1.71   & 0.55 &  8.80 & \num{1.5e5} \\
  \hline
  \hline
    \end{tabular*}
    \caption{Summary of the results presented in Fig.~\ref{fig:laassist}. Here, the minimum and maximum value of the dimensionless enhancement coefficient, the corresponding energy and relaxation rate $\Gamma$ to the Fermi level are displayed.}
    \label{tab:la}
\end{table*}

Overall, laser-assisted nuclear excitation proves to be slightly more efficient than quenching schemes for nuclear deexcitation. This is primarily due to the fact that the direct excitation rate  $\Gamma_{\rm ex}$ with the pulsed VUV laser \cite{thielking2023vacuum} is relatively low.
With the given laser parameters and by choosing the most advantageous initial defect state, nuclear excitation can be enhanced by more than three orders of magnitude with the given set of parameters. Laser-assisted quenching is similarly strong, also achieving an enhancement of more than three orders of magnitude. However, not all electronic states are suitable for efficient nuclear (de)excitation.
As a general feature, small coefficients result from electronic states with small bridge rates, related to small dipole moments entering the bridge matrix element.
The highest coefficients correspond to the cases where bridge rates are significantly larger.

Exceptions appear, for instance, in $^{229}$Th:LiSAF $1\bar{1}\bar{1}01$ where the defect at 8.8 eV achieves the highest (de)excitation strengths.
This is due to the small photon energy entering the denominator of the quenching rate $\Gamma_Q\sim\Gamma_{\mathrm{sp}}/\omega_p^3$ appearing in the numerator of expression~\eqref{eq:quench}.   This factor compensates for the differences in the bridge rates, providing larger coefficients at smaller energies.

Comparing the overall performance of the two crystals, we find that in $^{229}$Th:LiCAF, the structures $0\bar{2}001$ and $\bar{1}0\bar{1}01$ lead to a substantial improvement in nuclear excitation and deexcitation.
It is therefore beneficial to specifically grow these structures, 
as their formation can be effectively controlled by tuning the fluorine concentration, see discussion in Sec.~\ref{sec:Crystal}.
In contrast, for $^{229}$Th:LiSAF, all structures perform well, except for $1\bar{1}\bar{1}01$, which exhibits significantly lower efficiency. Fortunately, this structure is energetically not the most favorable charge compensation scheme considered here.

\subsection{Defect state energy variation}
\label{sec:vary}
%%%%%%%%%%%%%%%%%%%%%%%%%%%%%%%%%%%%%

In this Section, we address in more detail the role of the defect state energies for the EB rate. To this end, we introduce, in addition to the rigid energy shift $\Delta_r$ used in the previous Sections to fix the conduction band edge at the measured value, a variable shift applied to the set of states in between the unoccupied thorium localized states with the lowest and highest energy. The energies of these states are simultaneously varied around the isomer energy by adding or subtracting the same value.

Our analysis focuses on quenching schemes as they are particularly relevant for clock operation. We calculate $\Gamma_Q(\ket{m,o}\rightarrow \ket{g, d_i})$
using a fixed final state $\ket{d_i}$, but varying the energy of the latter throughout the full range of interest around the isomer energy $E_{\mathrm{iso}}=8.355$ eV. The energies of the other defect states from the set are shifted at the same time accordingly. This leads to peaks in the spectrum, as one by one, the defect states will come in resonance with the isomer energy, producing diverging terms in the intermediate state summation.

\begin{figure}
    \centering
    \includegraphics[width=1\linewidth]{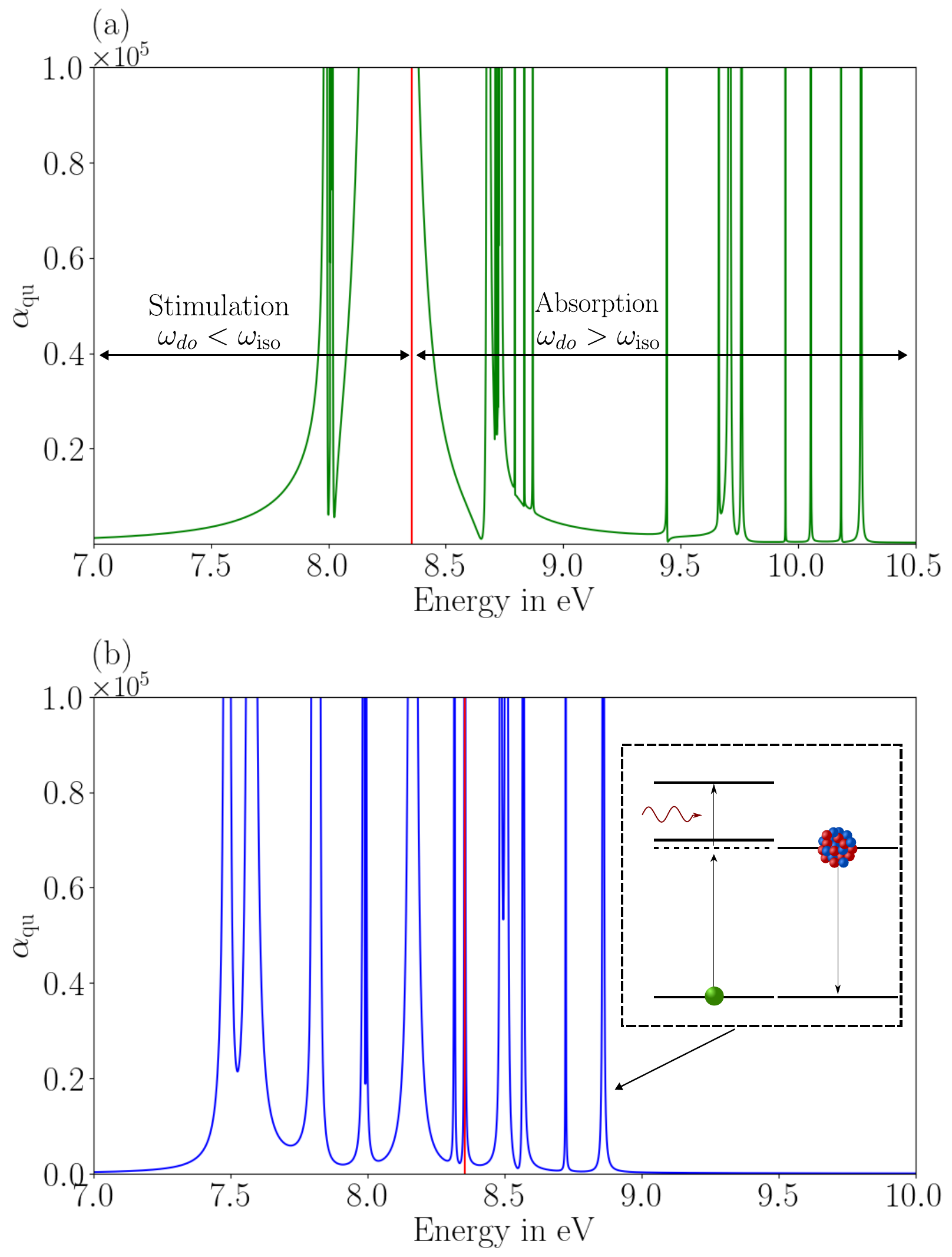}
    \caption{Quenching coefficient as a function of artificially varied energy of the final state $\ket{d_i}$ in (a) $^{229}$Th:LiCAF $\bar{1}0\bar{1}01$ and (b) $^{229}$Th:LiSAF $00\bar{1}11$. The red vertical lines mark the isomer energy $E_{\mathrm{iso}}=8.355$ eV. Left (right) thereof, the EB stimulation (absorption) scheme is used. For all calculations, we use $I_\omega= \SI{5}{\watt \second \per \metre \squared}$ and all other laser parameters as in the previous Section. }
    \label{fig:spectrum}
\end{figure}

Fig.~\ref{fig:spectrum} presents one such energy spectrum for each crystal. In both cases, the initial electronic state is considered to be $\ket{o}$.
Panel (a) displays a quenching spectrum for $^{229}$Th:LiCAF $\bar{1}0\bar{1}01$, where the final defect state $\ket{d_i}$ corresponds to the defect in Fig.~\ref{fig:sponti} with an energy of $E = 11.84$ eV  and has a norm of $\braket{\Psi|\Psi} = 0.71$. We vary its energy in the interval (7 eV, 10.5 eV). This range guarantees that the other defect states remain located within the band gap of the pristine host. 
Panel (b) shows the corresponding quenching spectrum for $^{229}$Th:LiSAF $00\bar{1}11$, with the final defect state $\ket{d_i}$ with energy  $E = 10.67$ eV in Fig.~\ref{fig:sponti} and $\braket{\Psi|\Psi} = 0.99$. Instead of using the calculated DFT value, the energy is varied now in the interval (7 eV, 10 eV).  
For both spectra, the required photon energies to quench the nuclear population are in the infrared to optical domain. We consider an assisting laser of intensity $I_\omega =\SI{5}{\watt \second \per \metre \squared}$.

We illustrate the position of the isomer energy value with the red vertical lines. Left of this line,  we have $\omega_{do}<\omega_{\mathrm{iso}}$ and quenching occurs via stimulation; to the right  $\omega_{do}>\omega_{\mathrm{iso}}$ and quenching is induced by absorption. The resonances visible in the spectrum correspond to an electronic state in the intermediate summation whose energy is becoming aligned with the energy of the nuclear isomer.
Physically, this means that a real electronic state comes close to the virtual state, as illustrated in the inset of Panel (b) for the case of absorption. This can lead to a large value of the quenching rate and hence $\alpha_{\mathrm{qu}}$. 

Depending on the true defect state energies, the correct EB quenching rate will be just one point (approximately) somewhere on the calculated lines in Fig.~\ref{fig:spectrum}. As expected, a real defect state close to the virtual one can strongly enhance the quenching rate. The exact enhancement is only approximated here for two reasons. First, we have neglected the imaginary part of the denominators in the EB matrix element \eqref{eq:bridge_matrix_element}, which avoids the divergence of the summation terms even when a real state would coincide with the considered virtual state. Second, once such a perfect resonance would occur, the additional photon is not required, and the occurring process is no longer EB but bound internal conversion or nuclear excitation by electron transition in the crystal. 
The experimental determination of the defect states' energies in the crystal is therefore paramount for determining the magnitude of EB quenching.

\section{Connection to experimental results}
%%%%%%%%%%%%%%%%%%%%%%%%%%%%%%%%%%%%%%%%%%%%%%%%%

\label{sec:experiment}
We begin this Section by summarizing intriguing experimental observations in three recent publications. These findings are then compared to our EB rate results.
\begin{enumerate}
    \item In Ref.~\cite{elwell2024laser} $^{229}$Th was excited in the  host crystal LiSAF. Two relevant spectroscopic features are reported. The first one was assigned to the radiative decay of the nuclear isomer. The second feature is a broad line at $\sim$ 6.81 eV which decays within a few seconds. This feature apparently stems from the 
    deexcitation of $^{229}$Th in the crystal environment. 
    
    Our DFT calculations do not indicate any electronic defect states below the nuclear isomer, such that a bridge decay remains (theoretically) energetically forbidden in the first place. Even by artificially shifting the DFT energies below the nuclear isomer, the value of the spontaneous rates would still remain orders of magnitude smaller than the observed short-term fluorescence.
    Hence, an EB decay seems highly unlikely in that case. It is therefore more likely that internal conversion via defect states caused this feature, as suggested in Ref.~\cite{elwell2024laser}. 
     
    \item In Ref.~\cite{terhune2024photo} laser assisted quenching was demonstrated in $^{229}$Th:LiSAF. Here, the VUV laser was shifted by 100 GHz from the thorium resonance after the excitation cycle. The authors then observed that the nuclear excited state population could be depleted. From these measurements, a quenching cross section was extracted. This cross section corresponds to a rate of $\Gamma \sim \mathcal{O}(\SI{6e2}{\per \second})$, which again seems too large to arise from EB based on our present results. Also, in this case, internal conversion from a defect in the band gap to the conduction band seems more plausible.
    
    \item In Ref.~\cite{schaden2024laser}, quenching at various laser wavelengths and powers, and sample temperatures was investigated in $^{229}$Th:CaF$_2$. In the following, we will focus on the room temperature results, because they are of greater reliability. Quenching strengths were measured 
    across a broad spectral range from VUV to IR. In particular, all lasers except the IR lasers could quench the nuclear population. While the VUV laser only slightly depleted the nuclear population, a maximum depletion emerged at UV energies of $4$ to $3$ eV, where quenching strengths up to 
    $\alpha_{\mathrm{qu}}\sim 3$ were reported at average cw laser powers of 30 mW.
    
    Although alternative mechanisms such as internal conversion or collective effects cannot be ruled out as the underlying causes, parallels can be drawn between the reported quenching strengths, older theoretical estimates for $^{229}$Th:CaF$_2$ in Ref.~\cite{bsnpra} and the order of magnitude estimates based on our EB calculations for LiCAF and LiSAF.
    Scaling the theoretical spectra in Fig.~7 in Ref.~\cite{bsnpra} to account for the experimentally used laser intensity roughly provides a compatible quenching rate. Similarly, despite using different crystals, by scaling the spectra in Fig.~\ref{fig:spectrum} with the factor $I_{\omega_2}/I_{\omega_1}$, where $I_{\omega_1}=\SI{5}{\watt \second \per \metre \squared}$ and $I_{\omega_2}=\SI{2.4e-4}{\watt\second\per\metre\squared}$ is the corresponding laser spectral intensity 
    for the experimentally reported quenching strengths $\alpha_{\mathrm{qu}}\sim 3$ in \cite{schaden2024laser}, quenching coefficients greater than one appear to be possible.
    EB quenching is therefore a plausible explanation for the experimental observations in Ref.~\cite{schaden2024laser}.
\end{enumerate}

% $I_{\omega_2}$ corresponds to the 420 nm laser with an average power of 30 mW, beam waist of \SI{250}{\micro \metre} and line width of $\Gamma_\ell= 2 \pi \times \SI{100}{\mega \hertz}$

\begin{figure}
    \centering
    \includegraphics[width=.8\linewidth]{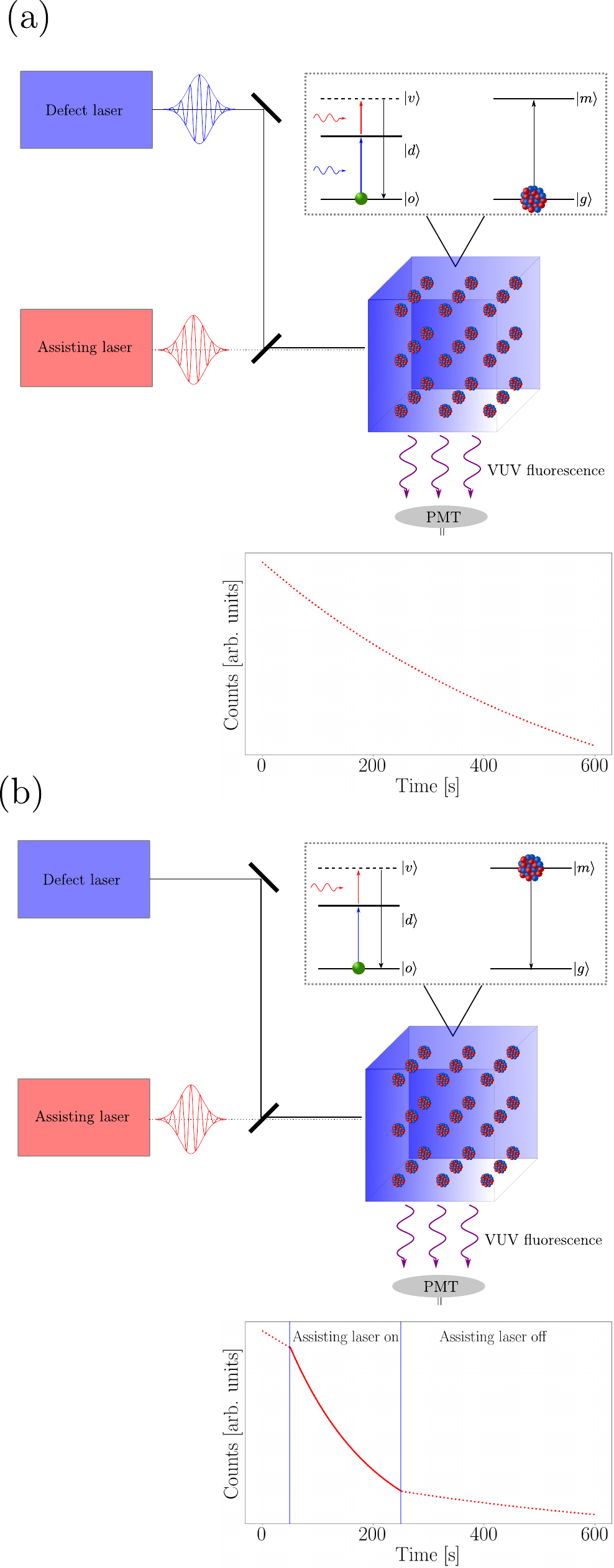}
    \caption{Quantum dynamical protocol to prove the existence of the EB decay channel for the scenario $\omega_{do}< \omega_{\mathrm{iso}}$. 
    (a) Excitation protocol for nuclear excitation. (b) Excitation protocol for nuclear deexcitation.}
    \label{fig:protocol}
\end{figure}

We conclude that to unambiguously identify the presence of an EB channel, more dedicated follow-up experimental and theoretical investigations are required.
A possible excitation protocol designed to support such a study is outlined in Fig.~\ref{fig:protocol}.
A somewhat similar proposal for Th ions was previously presented in Ref.~\cite{bilous2018laser}.
The main features of this protocol are two tuneable lasers, a $^{229}$Th-doped crystal, and photomultiplier tubes (PMTs) which can record VUV fluorescence.
The protocol begins with panel (a). In this step, two tunable lasers simultaneously interrogate the crystal. The first laser, referred to as the defect laser, is resonantly driving the transition from the electronic ground state to a defect state. The energy of this defect state must be determined in advance, either via VUV fluorescence spectroscopy, VUV absorption spectroscopy, or high precision electronic structure calculations \cite{thorstenaccurate}.
The second laser is called the assisting laser. Both lasers must satisfy the two-photon resonance condition
\begin{equation} 
\omega_d \pm \omega_a = \omega_{\mathrm{iso}}, 
\end{equation} 
where the $+$ sign corresponds to a defect level lying below the isomer, and the $-$ sign corresponds to a defect level lying above the isomer. Note that in Fig.~\ref{fig:protocol} only the case $\omega_{do} < \omega_{\mathrm{iso}}$ is displayed.

After the first interrogation cycle, both lasers are switched off, the shutters of the PMTs are opened, and VUV fluorescence is recorded. A lack of a signal may indicate that the assisting laser is either detuned from the bridge resonance or that its spectral power is insufficient, such that competing electron conversion decay channels prevent signal detection. This can be addressed by slightly adjusting the laser wavelength and/or increasing the laser power.
This procedure should be repeated until a clear signal is observed that can be attributed to the fluorescence of the nuclear isomer.
No fluorescence signal despite rigorous scanning might indicate that other decay processes within the crystal dominate, preventing the observation of a laser-assisted EB scheme.

In case of successful VUV photon detection, an approximate value for the EB rate can be extracted based on interrogation time and count rates.
This value is crucial for estimating the required spectral intensity for the second part of the protocol.
After the VUV fluorescence signal has subsided, almost all nuclei are in the ground state, and nuclear excitation is restarted via Fig.~\ref{fig:protocol} (a). However, instead of remeasuring the full VUV decay spectrum, the PMT shutters are closed, and one can proceed with the second part of the protocol, see Fig.~\ref{fig:protocol}(b). In this step, only the assisting laser irradiates the crystal containing the nuclei, which were excited in the first place via the procedure illustrated in Fig.~\ref{fig:protocol} (a). Based on the EB rate estimate from the full fluorescence signal, the laser's spectral intensity should be chosen such that $\alpha_{\mathrm{qu}} = \mathcal{O}(1)$.
Under these conditions, a characteristic drop in the nuclear excited-state population should be observed, similar to the laser-assisted quenching effects reported in \cite{schaden2024laser}. This behavior would provide evidence for the existence of an EB channel and allow for a determination of its strength from both the excitation and deexcitation processes.

This quenching technique could later be employed during clock operation to deplete the nuclear population on a much shorter timescale compared to its radiative decay. 
As discussed in Ref.~\cite{bsnpra},  this speed-up can be used to improve the short-term stability of a nuclear clock and reduce the time between interrogation cycles.

\section{Conclusion \& Outlook}

We have investigated EB schemes in $^{229}$Th:LiCAF and $^{229}$Th:LiSAF involving defect states.
These states, predicted by DFT, are located in or near the band gap, presumably in energetic proximity to the nuclear isomeric transition. Due to the structural complexity of both materials, only the four lowest charge compensation configurations were considered for each crystal. In $^{229}$Th:LiCAF, the energetically lowest configuration can be tuned via the chemical potential of flourine, while such control is not feasible for $^{229}$Th:LiSAF.

Our calculations for spontaneous EB schemes using the predicted DFT energies have also shown that delocalized defect states can contribute to large EB rates. The rate values are influenced by three factors: the emitted photon energy and the magnitudes of the dipole and hyperfine coupling matrix elements. It turns out that delocalized states with weak hyperfine coupling have, in turn, large dipole matrix elements, which can compensate for the former. 

When considering laser-assisted EB excitation and quenching schemes, we find, for both 
LiCAF and LiSAF, that enhancements up to three orders of magnitude are achievable compared to direct laser excitation or the spontaneous radiative decay of the isomer for a realistic set of parameters. These enhancements can be even stronger in case the defect state energies lie advantageously, and a real defect state is in close vicinity of the EB virtual state. However, the exact defect state positions are yet to be determined experimentally.

Finally, we have discussed our results in the context of recent experiments. While the calculated rates for spontaneous EB processes are much smaller than reported experimental observations, the quenching observations recently made in CaF$_2$ seem to roughly agree with theoretical EB quenching rate values. 
For a future unambiguous observation of EB, we propose an interrogation scheme that might provide direct evidence for EB excitation and quenching. Further theoretical and experimental studies will hopefully confirm the EB processes and demonstrate their use for future clock operation in VUV-transparent crystals.

%Since the EB process is not the only hyperfine-interaction-mediated process in the crystal environment, we plan to extend our investigations to electron conversion processes. These will focus on transitions from the valence band to the defect states, and from defect states to the conduction band, in various thorium doped insulators.
%We expect our results to be in a similar order of magnitude than the results presented \cite{morgan2024theory}.

\label{sec:end}
%%%%%%%%%%%%%%%%%%%%%%%%%%%%%%%%%%%%%%%%%%%%%%%%%%%%%%%%%%%%%%%%%%

\begin{acknowledgments}

%%%%%%%%%%%%%%%%%%%%%%%%%%%%%%%%%%%%%%%%%%%%%%%%%%%%%%%%%%%%%%%%%%%%%%%%%%
This research was supported by the Austrian Science Fund (FWF) [grant DOI:10.55776/F1004] (COMB.AT) together with the Deutsche Forschungsgemeinschaft (DFG, German Science Foundation) (PA 2508/5-1).
A.P. gratefully acknowledges support from the DFG in the framework of the Heisenberg Program (PA 2508/3-1).
This work has been funded by the European Research Council (ERC) under the European Union’s Horizon 2020 and Horizon Europe research and innovation programme (Grant Agreement No. 856415 and No. 101087184).
\end{acknowledgments}

\newpage
{
\onecolumngrid
\appendix

%%%%%%%%%%%%%%%%%%%%%%%%%%%%%%%%%
\section{Convergence}
\label{sec:appendix1}

The matrix element in Eq.~\eqref{eq:bridge_matrix_element} requires a summation over all unoccupied intermediate states $\ket{k}$. These involve both (de)localized defects and conduction band states %which do not correspond to the states $\ket{o}$, $\ket{d}$ involved in the bridge process. \tk{I removed this statement, since I found both ways in literature. There is no consequence for the results. Both approaches yield the same numbers with a very small deviation.}
In principle, the conduction band consists of infinitely many states. 
However, as shown in Ref.~\cite{bsnpra}, the EB rate should converge by including a large number of conduction band states in the intermediate summation.
We therefore briefly discuss here the convergence of our EB rates.

When the wave functions $\psi_n$ are extracted from VASP output, electronic states are assigned with a state index $n$ ascending in energy. We refer to the relevant states according to their index $n$. In the following, the Fermi level is labelled as $n=0$, while the first defect is labeled as $n=1$, and so on.

For illustration purposes, we consider the $^{229}$Th:LiSAF structure $00\bar{1}11$ and present convergence studies for three different cases.
Figure~\ref{fig:convergence} shows the EB rate for spontaneous nuclear excitation considering three initial defect states, each with a different degree of localization. The EB rate is plotted
as a function of the number of included electronic states $\ket{k}$ in the intermediate state summation.
In Fig.~\ref{fig:convergence}, (a) corresponds to the initial localized defect state with index $n=6$ at $\sim$ \SI{10.69}{\electronvolt}, (b) to the semi-localized defect state with index $n=3$ at $\sim$ \SI{10.46}{\electronvolt} and (c) to the delocalized state with index $n=12$ at $\sim$ \SI{11.22}{\electronvolt}. 
\begin{figure}[H]
    \centering
    \includegraphics[width=1\linewidth]{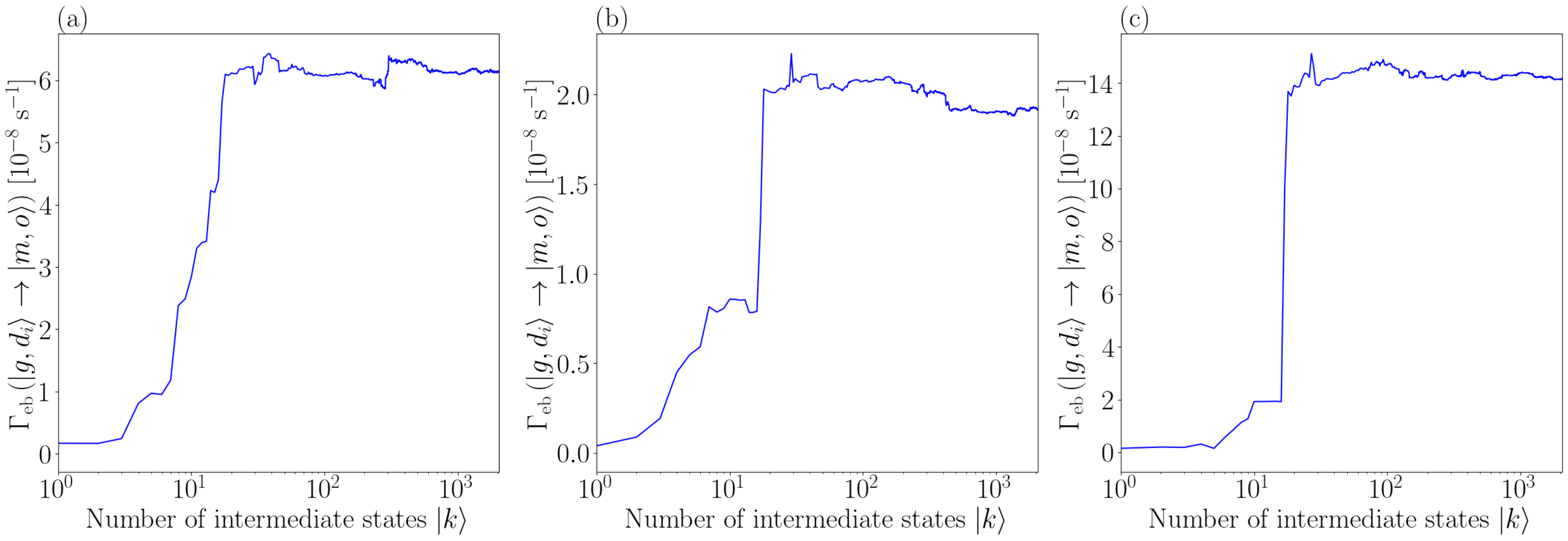}
    \caption{Spontaneous EB rate as a function of number of intermediate electronic states $\ket{k}$ for the $^{229}$Th:LiSAF structure $00\bar{1}11$. The considered initial defect state is (a) $n=6$, (b)  $n=3$ and (c)  $n=12$. See text for further explanations.}
    \label{fig:convergence}
\end{figure}

Overall, more than 2000 intermediate states are included in the summation.  In this logarithmic representation, we observe that for a small number of states, the rate grows very fast. This happens when the unoccupied defect states enter the intermediate state summation. These states contribute the most to the rate as their energies are close to the nuclear isomer and their hyperfine coupling is stronger than for delocalized states, as discussed in Appendix \ref{appendix}. Already after a few hundred intermediate states, a stable plateau in the EB rate is reached for all three studied cases.

%%%%%%%%%%%%%%%%%%%%%%%%
\section{Dipole moments and hyperfine couplings of $^{229}$Th:LiCAF $0\bar{2}001$}
\label{appendix}

As discussed in the main text, there is a large spread in the order of magnitude of the spontaneous EB rates ranging from $\mathcal{O}(\SI{e-13}{\per \second})$ to $\mathcal{O}(\SI{e-6}{\per \second})$. In the following, we will discuss this range based on the example of the $^{229}$Th:LiCAF structure $0\bar{2}001$, see Fig.~\ref{fig:sponti} (a),  where the minimum value of the rate $\Gamma_{\mathrm{eb}}\sim\mathcal{O}(\SI{e-13}{\per \second})$ originates from considering the defect state $n=2$ at $\sim 8.8$ eV as initial state. The maximum EB rate $\Gamma_{\mathrm{eb}}\sim\mathcal{O}(\SI{2e-7}{\per \second})$ was obtained considering the semi-localized state $n=11$ at $\sim 11.1$ eV as initial state.

First, we investigate the dipole moment matrix elements, which were extracted from VASP. For this purpose, we define
\begin{equation}
 \mathcal{Q}^2_{fi} =   \ab{\bra{f} \bm{\mathcal{Q}}_{E1} \ket{i}}^2\, .
\end{equation}
Figure~\ref{fig:dipolematrix} presents the corresponding matrix, with each pixel color illustrating the matrix element  $\mathcal{Q}^2_{fi}$ value. The abscissa (ordinate) give the initial (final) state $\ket{i}$  ($\ket{f}$)  used in the calculation, labeled according to the state number as explained in Appendix~\ref{sec:appendix1}.
\begin{figure}[]
    \centering
\includegraphics[width=0.6\linewidth]{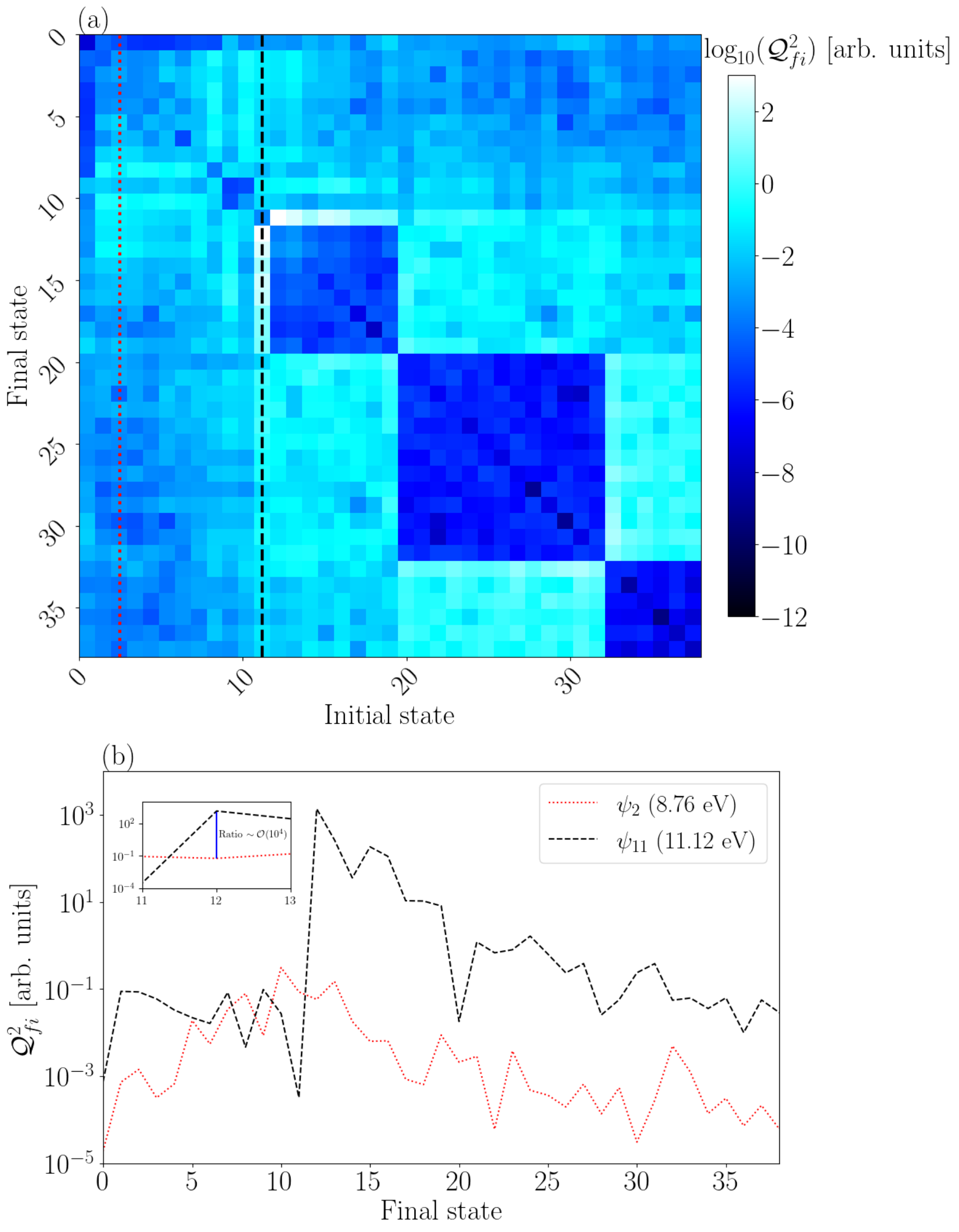}
    \caption{(a) Dipole matrix (in arbitrary units) for $^{229}$Th:LiCAF $0\bar{2}001$. (b) Dipole matrix elements modulus squared as a function of final state for $n=2$ (red dotted) and $n=11$ (black dashed) as initial states. }
    \label{fig:dipolematrix}
\end{figure}
In this matrix, the red dotted line displays matrix elements with $n=2$ as initial state, and the black dashed lines correspond to $n=11$ as initial state.
We observe that the dipole moments in $n=11$ are, in most cases, larger than the dipole moments involving the band $n=2$.
In particular, as shown in Fig.~\ref{fig:dipolematrix} (b), the projection along the marker line shows that the matrix elements differ most for the final state $n=12$, with values stretching over approx. 4 orders of magnitude. 

Next, we investigate the influence of the hyperfine operators on the EB matrix element. For this purpose, we recall the expression of the EB matrix element modulus squared:
\begin{equation}
    \ab{\bra{m,f} \Tilde{\bm{\mathcal{Q}}}_{E1} \ket{g,i}}^2=\ab{ \sum_{\lambda K, q} (-1)^q \Big[\sum_k \frac{\bra{f}\bm{\mathcal{Q}}_{E1}\ket{k}\bra{k}\mathcal{T}_{\lambda K, q}\ket{i}}{\omega_{dk}- \omega_{\text{iso}}} 
        +  \frac{\bra{f}\mathcal{T}_{\lambda K,q}\ket{k}\bra{k}\bm{\mathcal{Q}}_{E1}\ket{i}}{\omega_{ok}+\omega_{\text{iso}}}\Big]
        \times \bra{m}\mathcal{M}_{\lambda K, -q}\ket{g}}^2\, .
\end{equation}

For an order of magnitude estimate, we consider, for simplicity, that 
\begin{equation}
    \frac{1}{\omega_{dk}-\omega_{\text{iso}}} = \frac{1}{\omega_{ok}-\omega_{\text{iso}}} = 1 \quad \& \quad \bra{f}\bm{\mathcal{Q}}_{E1}\ket{k}=\bra{k}\bm{\mathcal{Q}}_{E1}\ket{i} = \bm{\chi}_k \quad \& \quad \bra{m}\mathcal{M}_{\lambda K, -q}\ket{g}=1
\end{equation}
where $\bm{\chi}_n$ is an arbitrary unit vector satisfying $\bm{\chi}_k\cdot\bm{\chi}_p = \delta_{k,p}$.
The remaining hyperfine operator matrix elements, which we term here effective hyperfine strength, read
\begin{equation}
    \mathcal{T}^2_{fi} = \sum_{\lambda K}\ab{\sum_{q,k}(-1)^q \lrb{\bra{k}\mathcal{T}_{\lambda K,q}\ket{i}+\bra{f}\mathcal{T}_{\lambda K,q}\ket{k}}}^2\, ,
\end{equation}
where the summation runs again over more than 2000 conduction band states.
In Fig.~\ref{fig:hyperfine}, we present the quantity $T_{fi}^2$ as a function of the initial band $\ket{i}$.
\begin{figure}[]
    \centering
\includegraphics[width=0.7\linewidth]{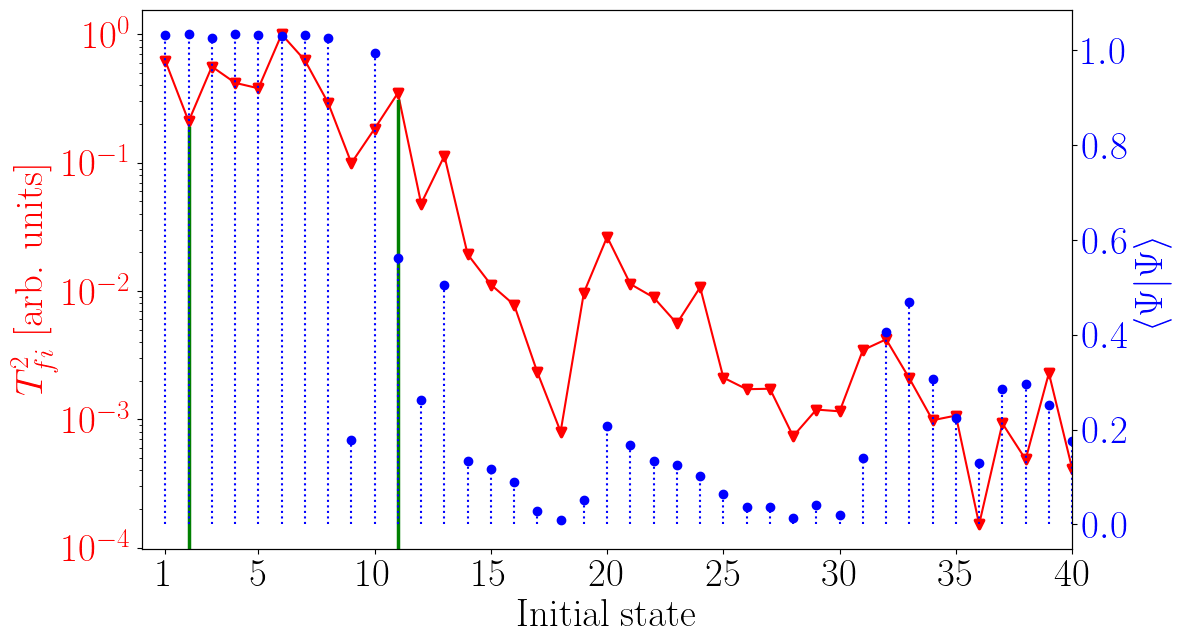}
    \caption{Effective hyperfine strength (in arbitrary units) as a function of initial state. In all cases, the Fermi level $n=0$ is the final state. Green vertical lines mark the relevant states $n=2$ and $n=11$.}
    \label{fig:hyperfine}
\end{figure}
As a general trend, the larger the degree of localization, the larger the hyperfine matrix element contribution. Exceptions, for instance for $n=20$, occur when large hyperfine matrix elements between localized and delocalized states are included in the summation over intermediate states. The case of the initial states $n=2$ and $n=11$ for which the total EB rate reaches a minimum or maximum is highlighted with the green vertical lines. Surprisingly, it turns out that the effective hyperfine strength $T_{fi}^2$ has the same order of magnitude for both initial states. 
Hence, we conclude that the six orders of magnitude between the minimum and maximum EB rates stem from the combination of the differences in photon energy $\omega^3_{p_1}/\omega^3_{p_2} \sim\mathcal{O}(\num{e2})$ and the difference in dipole matrix elements $\mathcal{Q}^2_{1}/\mathcal{Q}^2_{2}\sim\mathcal{O}(\num{e4})$.

\newpage

\newpage

\twocolumngrid
 
\bibliography{refs}

\end{document}